\documentclass[letterpaper,11pt]{article}

\usepackage[backend=bibtex]{biblatex}
\usepackage[margin=1in]{geometry}
\usepackage[T1]{fontenc}
\usepackage{lmodern}

\usepackage[pdftex,hyperfootnotes=false]{hyperref} 
\hypersetup{colorlinks=false}

\usepackage{url}
\usepackage{booktabs}
\usepackage{amsfonts}
\usepackage{nicefrac}
\usepackage{microtype}
\usepackage{lipsum}
\usepackage{graphicx}
\usepackage{framed,multirow}
\usepackage{amssymb}
\usepackage{mathrsfs}

\usepackage{latexsym}
\usepackage{bm}

\graphicspath{{images/}}

\usepackage{url}
\usepackage{xcolor}
\usepackage{amsmath}
\usepackage{amsthm}
\usepackage{parskip}
\usepackage{fancyhdr}
\usepackage{xcolor}
\usepackage{pgf}
\usepackage[font=small,labelfont=bf]{caption}
\usepackage{subcaption}
\usepackage{epstopdf}
\usepackage{pgfplots}
\usepackage{color}
\usepackage{tikz}
\usetikzlibrary{shapes.geometric}
\usepackage{float}
\usepackage{bm}
\usepackage{stmaryrd}
\usepackage{cleveref}

\usetikzlibrary{arrows, arrows.meta}
\definecolor{newcolor}{rgb}{.8,.349,.1}
\usetikzlibrary{calc}
\usetikzlibrary{external}
\usepackage{calrsfs}
\usepackage{enumitem}
\pgfplotsset{compat=1.15} 
\usepackage{scrextend}
\usepackage{algorithm}
\usepackage{algcompatible}
\usepackage{algpseudocode}
\usepackage[justification=centering]{caption}
\usepackage[bottom]{footmisc}

\let\oldparagraph=\paragraph
\renewcommand\paragraph[1]{\oldparagraph{#1.}}

\def\be{\begin{equation}}
\def\ee{\end{equation}}
\def\ba{\begin{array}}
\def\ea{\end{array}}
\def\bea{\begin{eqnarray}}
\def\eea{\end{eqnarray}}
\def\beas{\begin{eqnarray*}}
\def\eeas{\end{eqnarray*}}

\newcommand{\Hcurl}{H({\rm curl})}

\newcommand{\Hdiv}{H({\rm div})}
\newcommand{\bfHdiv}{\bfH({\rm div},\Omega)}
\newcommand{\bfHdivK}{\bfH({\rm div},K)}
\newcommand{\bfHdivh}{\bfH({\rm div},\Omega_h)}

\newcommand{\bfH}{\boldsymbol{H}}

\newcommand{\ptl}{{\partial}}

\newcommand{\ds}{\displaystyle}

\catcode `@=11

\newbox \itemlist@label
\newdimen \itemlist@labelpad

\def\itemlist@makelabel#1{%
\setbox\itemlist@label =\hbox{#1}%
\ifdim \wd\itemlist@label >\labelwidth
\itemlist@labelpad=\textwidth
\advance\itemlist@labelpad by -\rightmargin
\advance\itemlist@labelpad by -\@totalleftmargin
\advance\itemlist@labelpad by \labelwidth
\hbox to \itemlist@labelpad {#1\hfill}%
\else #1\hfill
\fi
}

\makeindex

\newtheoremstyle{problemstyle}  
{3pt}                                               
{3pt}                                               
{\normalfont}                               
{}                                                  
{\bfseries}                 
{\normalfont\bfseries:}         
{.5em}                                          
{}                                                  
\theoremstyle{problemstyle}

\newtheoremstyle{Assumptionstyle}  
{3pt}                                               
{3pt}                                               
{\normalfont}                               
{}                                                  
{\bfseries}                 
{\normalfont\bfseries:}         
{.5em}                                          
{}                                                  
\theoremstyle{Assumptionstyle}

\newcommand{\vertiii}[1]{{\left\vert\kern-0.25ex\left\vert\kern-0.25ex\left\vert #1 
    \right\vert\kern-0.25ex\right\vert\kern-0.25ex\right\vert}}

\title{
\LARGE
An Anisotropic \texorpdfstring{$hp$}{hp}-Adaptation Framework\\[8pt] for Ultraweak Discontinuous Petrov--Galerkin Formulations
}

\author{
{
\Large 
Ankit Chakraborty\footnote{Corresponding author: ankit.chakraborty@austin.utexas.edu},
Stefan Henneking, Leszek Demkowicz
}\\[5pt]
Oden Institute, The University of Texas at Austin
}
\date{\today}

\begin{document}
\clearpage \maketitle

\textbf{Abstract:}
In this article, we present a three-dimensional anisotropic $hp$-mesh refinement strategy for ultraweak discontinuous Petrov--Galerkin  (DPG) formulations with optimal test functions. The refinement strategy utilizes the built-in residual-based error estimator accompanying the DPG discretization. The refinement strategy is a two-step process: (a) use the built-in error estimator to mark and isotropically $hp$-refine elements of the (coarse) mesh to generate a finer mesh; (b) use the reference solution on the finer mesh to compute optimal $h$- and $p$-refinements of the selected elements in the coarse mesh. The process is repeated with coarse and fine mesh being generated in every adaptation cycle, until a prescribed error tolerance is achieved. We demonstrate the performance of the proposed refinement strategy using several numerical examples on hexahedral meshes.

\vspace{10pt}

\textbf{Acknowledgments:} 
We thank Jacob Badger for fruitful discussions. Ankit Chakraborty, Stefan Henneking and Leszek Demkowicz were supported with NSF award 2103524. Ankit Chakraborty is partially supported with the Peter O’Donnell Jr.~Postdoctoral Fellowship. All numerical experiments in this article were performed on {\em Frontera}'s Intel Cascade Lake (CLX) nodes located at the Texas Advanced Computing Center \cite{frontera} using $\text{DMS22025}$ allocation.

%
%

\section{Introduction}

Automatic $hp$-mesh refinement algorithms are powerful tools that aid finite element discretizations in computing solutions of partial differential equations (PDEs) in an efficient and accurate manner. They achieve this efficiency and accuracy by constructing meshes with optimally distributed element size
$h$ and polynomial order of approximation $p$ \cite{LDbook_a,LDbook_b}. Finite element meshes with optimal element size and polynomial distribution are critical for resolving solution features such as boundary layers in convection-dominated diffusion problems or point and edge singularities in problems with re-entrant corners. In such problems, optimal $hp$-meshes are indispensable for achieving exponential convergence \cite{babuska_a,babuska_b,wr_a,schw_a,babuska_c}. Designing algorithms capable of generating a sequence of optimal $hp$-meshes that deliver optimal convergence rates in a problem-agnostic manner has been a significant challenge in finite element research over the past few decades \cite{wr_a,hp_dem_a,hp_dem_b,hp_dem_c}. Typically, automatic mesh refinement strategies are driven by computable error estimates. These error estimates are computed using the approximate solution delivered by the discretization scheme. Therefore, the accuracy and stability of the underlying numerical discretization are paramount for the effectiveness of the mesh refinement strategy.

The discontinuous Petrov--Galerkin (DPG) method with optimal test functions, first introduced by Demkowicz and Gopalakrishnan in \cite{LD_a,LD_b,LD_c}, has emerged as a critical technology in terms of robustness and stability over the past decade. Given a stable variational formulation of an underlying PDE and a trial approximation space, the DPG method computes a test space so that the resulting discretization is {\em inf--sup} stable. The methodology delivers an orthogonal projection in the so-called energy norm.
Another significant advantage of the DPG methodology is the presence of a built-in residual-based error estimator, also known as the energy error estimate. This makes the DPG method an ideal candidate for automatic mesh optimization algorithms.

In this article, we focus on the ultraweak DPG finite element formulation with optimal test functions and propose a problem-agnostic anisotropic $hp$-mesh refinement strategy. It is critical to mention that, for the ultraweak DPG method, the energy norm is {\em equivalent} to the $L^2$-error \cite{LD_e}. Consequently, the method delivers essentially the $L^2$-projection of the unknown solution.

The proposed refinement strategy consists of  the following steps:
\begin{itemize}
\item \textbf{Step 1}: Solve the problem on the current {\em coarse mesh}.
\item \textbf{Step 2}: Utilize the computed DPG residual to mark coarse mesh elements for refinements.
\item \textbf{Step 3}: Isotropically $hp$-refine the marked elements to generate a {\em fine mesh}. 
\item \textbf{Step 4}: Solve the problem on the fine mesh to obtain the fine mesh solution $u$.
\item \textbf{Step 5}: Use the fine mesh solution $u$ as a {\em reference solution} to determine optimal (anisotropic) $hp$-refinements of the selected coarse grid elements.
\item \textbf{Step 6}: Restore the coarse mesh and execute the optimal $hp$-refinements.
\end{itemize}
We essentially use the $hp$-algorithm from \cite{LDbook_a,LDbook_b}. 
Optimizing the mesh in the $L^2$-space greatly simplifies the original procedure. There is no need for mesh optimization on edges and faces; the {\em Projection-Based Interpolation} reduces to the $L^2$-projection performed on elements only.
The optimal refinements of a coarse element $K$ are determined by maximizing the rate $(e_{hp})$ with which the projection error decreases,
\[
e_{hp} := \frac{\Vert u - P_\mathrm{coarse} u \Vert^2 - \Vert u - P_\mathrm{opt} u \Vert^2}{N_\mathrm{opt} - N_\mathrm{coarse}}\, .
\]
Here, $P_\mathrm{coarse}$ denotes the $L^2$-projection onto the coarse mesh, $P_\mathrm{opt}$ is the projection onto the optimal mesh to be determined, 
$N_\mathrm{opt}$ and $N_\mathrm{coarse}$  denote the number of degrees-of-freedom (dof) of the optimal and coarse grid elements, respectively. As the $L^2$-projection onto discontinuous polynomial spaces is a purely local operation, the mesh optimization can be trivially performed in parallel.

The article is organized as follows. Section~\ref{sec:DPG_discretization} briefly introduces the ultraweak DPG finite element discretization with optimal test functions. Section~\ref{sec:optimization} provides the details of the mesh optimization algorithm. In Section~\ref{sec:results}, numerical results demonstrate the efficacy of the proposed refinement strategy. Finally, we conclude with a short discussion in Section~\ref{sec:conclusions}.

%
%

\section{DPG Methodology
\label{sec:DPG_discretization}
}

The core idea behind the (ideal) DPG method is to automatically generate a stable discretization for a given well-posed variational formulation and an approximate trial space. The method achieves stability by computing an optimal discrete test space \cite{LD_b} corresponding to the approximate trial space in such a way that the supremum over the continuous test space in the discrete {\em inf--sup} \cite{Babuska1971} is automatically attained over the discrete test space. The optimal test space is obtained by inverting the Riesz map 
corresponding to the test inner product over a {\em discontinuous or broken}\footnote{Hence the ``D'' in the DPG method.} test space. Unfortunately, inverting the Riesz operator exactly is impossible due to the infinite-dimensional nature of the continuous test space. Thus, in practical realizations of DPG methods, we approximate the inverse of the Riesz operator by inverting the Gram matrix induced by the test norm on a larger, but finite-dimensional {\em enriched} discontinuous test space.\footnote{We then refer to it as the {\em practical} DPG method.} The use of broken test spaces enables element-wise inversion of the Gram matrix, but it also introduces trace variables defined on the mesh skeleton \cite{carstensen_a}. 

We consider a model Poisson problem. Let $\Omega \subset \mathbb{R}^3$ be a bounded Lipschitz domain with boundary $\Gamma$ split into two disjoint parts: $\Gamma_u$ and $\Gamma_{\sigma}$. The first-order formulation of the Poisson problem is given by:

\begin{equation}
\arraycolsep=1.4pt
\left\{
\begin{array}{rllll}
 \bm{\sigma} - \nabla u&= \bm{0} \quad &\text{in} \quad \Omega, \\ 
 -\nabla \cdot \bm{\sigma} & = f \quad &\text{in} \quad \Omega,  \\    
 u & = u_0 \quad &\text{on} \quad \Gamma_u, \\
 \bm{\sigma} \cdot \bm{n} & = \sigma_0 \quad &\text{on} \quad \Gamma_\sigma,  \\
 \end{array}
 \right.
\end{equation}

where $f \in L^2(\Omega)$ represents the source term and $\bm{n}$ denotes the outward normal. Before presenting the ultraweak variational formulation, we briefly introduce the energy spaces used in this article. We define the standard energy spaces as:
\begin{align}
L^2(\Omega) &= \left\{ u : \Omega \rightarrow \mathbb{R}: {\Vert u \Vert} < \infty \right\}, \nonumber \\
H^1(\Omega) &= \left\{ v : \Omega \rightarrow \mathbb{R}:  v \, \in \, L^2(\Omega), \nabla v \, \in \, {\left(L^2(\Omega)\right)}^3   \right\}, \\
\bfHdiv
&= \left\{  \bm{w}: \Omega \rightarrow \mathbb{R}^3: \bm{w} \, \in \, (L^2(\Omega))^3, \nabla \cdot \bm{w} \, \in \, L^2(\Omega)	\right\}. \nonumber
\end{align}
In the DPG method, discontinuous energy spaces are used for the test functions. Thus, we must define broken equivalents of $H^1(\Omega)$ and $\bfHdiv$ spaces for the finite element mesh ($\Omega_h$):
\begin{align}
\begin{split}
H^1(\Omega_h) &:= \left\{ v : \Omega \rightarrow \mathbb{R}: v\big|_{K} \in H^1(K) \quad \forall \, K \, \in \, \Omega_h  \right\}, \\
\bfHdivh &:=  \left\{ \bm{w} : \Omega \rightarrow \mathbb{R}^3: \bm{w}\big|_{K} \in \bfHdivK \quad \forall \, K \, \in \, \Omega_h 	\right\},
\end{split}
\end{align}
where $K \in \Omega_h$ represents an element of the finite element mesh. Use of the broken test spaces \cite{carstensen_a} leads to the introduction of additional trace unknowns on the mesh skeleton. The traces spaces are defined as:
\begin{align}
\begin{split}
H^{1/2}(\Gamma_h) &:= \left\{ \hat{u} : \exists \, u \, \in \, H^1(\Omega) \, \text{such that} \, \hat{u} = \gamma^K(u\big|_{K}) \, \text{on} \, \partial K \quad \forall \, K \in \Omega_h  \right\}, \\
H^{-1/2}(\Gamma_h) &:= \left\{ \hat{\sigma}_n : \exists \, \bm{\sigma} \, \in \bfHdiv \, \text{such that} \, \hat{\sigma}_n = \gamma_n^K(\bm{\sigma} \big|_{K}) \, \text{on} \, \partial K \quad \forall \, K \in \Omega_h  \right\},
\end{split}
\end{align}
where  $\gamma^K$ and $\gamma_n^K$ represent continuous and normal trace operators, respectively \cite{lec_LD}.

\paragraph{Ultraweak formulation}  
Let $(U,\hat{U})$ be the approximation trial space, $V$ the test space, and $V^{'}$ the dual space of $V$.  Then, the ultraweak DPG formulation of the Poisson problem can be stated as: Given $l \in V^{'}$, find $\mathfrak{u} \, \in \, U$ and $\hat{\mathfrak{u}} \in \hat{U}$ satisfying:
\be
b(\mathfrak{u},\mathfrak{v}) + \hat{b}(\hat{\mathfrak{u}},\mathfrak{v}) = l(\mathfrak{v}) \quad \forall \, \mathfrak{v} \in  V,
\ee
where
\be \label{UW_disc}
\begin{split}
\mathfrak{u} &= (u,\bm{\sigma}) \, \in \, L^2(\Omega) \times (L^2(\Omega))^3, \\
\hat{\mathfrak{u}} &= (\hat{u},\hat{\sigma}_n) \, \in \, H^{1/2}(\Gamma_h) \times H^{-1/2}(\Gamma_h): \hat{u} = u_0 \, \text{on} \, \Gamma_u, \hat{\sigma}_n = \sigma_0 \, \text{on} \, \Gamma_\sigma, \\
\mathfrak{v} &= (v,\bm{\tau}) \, \in \, H^1(\Omega_h) \times \bfHdivh, \\
b(\mathfrak{u},\mathfrak{v}) &= (\bm{\sigma},\nabla v)_{\Omega_h} + (\bm{\sigma},\bm{\tau})_{\Omega_h} + (u,\nabla \cdot \bm{\tau})_{\Omega_h}, \\
\hat{b}(\hat{\mathfrak{u}},\mathfrak{v}) &= -{\left\langle \hat{u}, \bm{\tau}\cdot \bm{n} \right\rangle}_{\Gamma_h} - {\left\langle \hat{\sigma}_n,v \right\rangle}_{\Gamma_h},\\
l(\mathfrak{v}) &= (f,v)_{\Omega_h} .
\end{split}
\ee
In~\ref{UW_disc}, $\left\langle \cdot , \cdot \right\rangle_{\Gamma_h}$ represents duality pairings defined over mesh skeleton $\Gamma_h$,
\be
\begin{split}
\langle \hat{u}, \bm{\tau}\cdot \bm{n} \rangle_{\Gamma_h} &\ds := \sum_{K \in \Omega_h}  \langle \hat{u} , \bm{\tau}\cdot \bm{n}_K \rangle_{\ptl K}\, , \\[2pt]
\langle \hat{\sigma}_h,v \rangle_{\Gamma_h} & \ds := \sum_{K \in \Omega_h}  \langle \hat{\sigma}_h,v \rangle_{\ptl K}\, ,
\end{split}
\ee
and
\be
{(\cdot,\cdot)}_{\Omega_h} = \sum_{K \, \in \, \Omega_h} (\cdot,\cdot)_{L^2(K)} \, .
\ee
The broken test space is equipped with the adjoint graph norm\cite{adj_norm_a,adj_norm_b}:
\be
{\Vert \mathfrak{v} \Vert}^2_V := {\Vert {A}_h^{\star} \mathfrak{v} \Vert}^2 + \alpha {\Vert \mathfrak{v} \Vert}^2
\ee
where $\alpha > 0$ is a scaling constant, and ${A}_h^{\star} \mathfrak{v} = {(\nabla \cdot \bm{\tau}, \nabla v + \bm{\tau})}_{\Omega_h}$ is the (formal) adjoint operator of $A_h \mathfrak{u} = {(\bm{\sigma} - \nabla u, -\nabla \cdot \bm{\sigma})}_{\Omega_h}$ computed element-wise. In this paper, all numerical experiments use $\alpha = 1$. Next, we briefly discuss the built-in error estimator.  Let $V_h(K) \subset V(K)$ be the {\em enriched} finite-dimensional  test space approximating the element test space $V(K)$, and $(U_h,\hat{U}_h) \subset (U,\hat{U})$ the finite-dimensional approximate trial space.   The basis functions for $V_h(K)$,  $U_h$ and $\hat{U}_h$ are denoted by $\varphi_i$,  $\psi_i$ and $\hat{\psi}_i$ respectively.  From~\ref{UW_disc},  we construct the following matrices for an element $K \in \Omega_h$,
\be
\begin{split}
{\mathrm{G}}_{K,lj} &= {(\varphi_l,\varphi_j)}_{V} \, , \\[8pt]
\mathrm{B}_{K,ij} &= b_K(\varphi_i,\psi_j) \, , \\[8pt]
\mathrm{\hat{B}}_{k,ij} &= \hat{b}_K(\varphi_i,\hat{\psi}_j) \, , \\[8pt]
\mathrm{l}_{K,i} &= l_K(\varphi_i),
\end{split}
\ee
where ${\mathrm{G}}_{K,lj}$ represents the element Gram matrix corresponding to the test inner product, $\mathrm{B}_{K,ij} $ represents the element stiffness matrix corresponding to the $L^2$ variables,  $\mathrm{\hat{B}}_{k,ij}$ represents the element stiffness matrix corresponding to the trace variables,  and $\mathrm{l}_{K,i}$ is the
element load vector. As usual,
$b_K(\cdot,\cdot)$,  $\hat{b}_K(\cdot,\cdot)$ and $l_K(\cdot)$ denote element $K$ contributions to bilinear forms $b(\mathfrak u,\mathfrak v), \hat{b}(\hat{\mathfrak{u}}, \mathfrak v)$, and linear form
$l(\mathfrak v)$, respectively.
An in-depth exposition of the algebraic structure of the linear system induced by DPG formulation for a diffusion problem can be found in \cite{LD_b, LD_d}.  

The built-in energy error estimate for a mesh element $K$ in the finite element mesh $(\Omega_h)$ is given by:
\be
	{\Vert (\mathfrak u, \hat{\mathfrak{u}}) - (\mathfrak u_h, \hat{\mathfrak{u}}_h) \Vert}_{E,K}^2 := 
	{\Vert {{R}_{{V}}}^{-1}  \left(  l_K(\cdot ) - b_K (\mathfrak u_h, \cdot ) - \hat{b}_K (\hat{\mathfrak{u}}_h, \cdot ) \right) \Vert}_{V(K)}^2
\ee
where 
\be
	R_V \, : V(K) \to (V(K))^\prime
\ee
is the Riesz operator corresponding to the test inner product.  With the element test space $V(K)$ approximated by a finite-dimensional enriched subspace $V_h(K)$, the element error indicators are computed as:
\be
{\eta_K}:= {\Vert {\mathrm{G}}^{-1} (\mathrm{l}_K - \mathrm{B}_K \mathrm{u_h} - \mathrm{\hat{B}}_K \mathrm{\hat{u}_h} ) \Vert}^2_{V(K)}\, .
\label{eq:element_error_indicator}
\ee

%
%

\section{Determining Optimal \texorpdfstring{$\boldmath{hp}$}{hp} Refinements
\label{sec:optimization}
}

The $hp$-algorithm described in this section is {\em exactly} the algorithm from \cite{LDbook_a,LDbook_b}, but specialized to the $L^2$-energy space. The corresponding algorithms for the $H^1$, $\Hcurl$, and $\Hdiv$ energy spaces, all based on minimizing the {\em Projection-Based (PB) interpolation error}, are significantly more intricate and consist of several steps reflecting the nature of the particular energy space. For instance, the algorithms for $H^1$ and $\Hcurl$ spaces consist of three stages involving mesh optimization on (interiors of) edges, faces and, finally, elements. The optimal mesh determined in each step serves as a starting point for the optimization in the subsequent step.

\begin{figure}[!htb]
\centering
\includegraphics[width=0.5\textwidth]{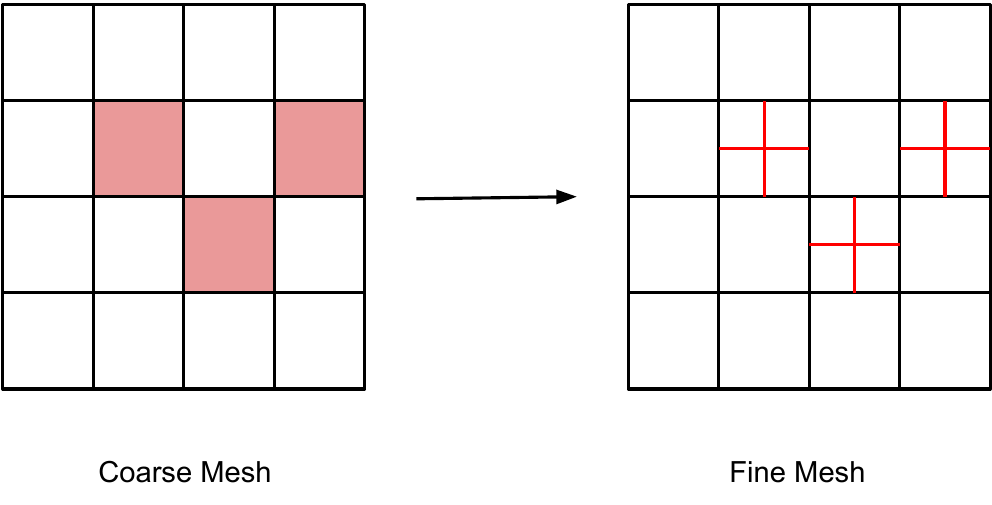}
\caption{Isotropic $hp$-refinement of the marked elements: the elements marked for refinement are shaded in red on the coarse mesh.} \label{fine_mesh}
\end{figure}

In the case of the $L^2$-energy space, there are no global conformity requirements; the PB interpolation reduces to just the $L^2$-projection, and the mesh optimization takes place over elements only.  The implementation of the algorithm is thus much simpler. The second difference between the presented and the original $hp$-algorithm lies in the involved elements. In the original algorithm, the optimization takes place over {\em all elements}, whereas here it only does for elements marked for refinement by the DPG residual. The number of elements entering the mesh optimization is thus much smaller.\footnote{Dependent upon the parameter in the D\"{o}rfler strategy \cite{dorflerstrat}.} The {\em fine mesh} providing the {\em reference solution} for the mesh optimization is also much smaller than the globally $hp$-refined mesh used in \cite{LDbook_a,LDbook_b}.~\Cref{fine_mesh} illustrates a two-dimensional case of mesh elements being marked by the DPG residual, followed by their isotropic $hp$-refinement\footnote{For a three-dimensional hexahedral element, isotropic $hp$-refinement denotes an isotropic $h_8$-refinement followed by an isotropic $p$-refinement of order 1.} to generate the {\em fine mesh}. 

The $hp$-algorithm consists of three steps: the first and third step are purely local (can be done trivially in parallel) while the second step requires a global reduction over the elements preselected for refinement by the DPG residual.

\subsection{Step 1: Staging a Competition of Refinements}

In the first step of the algorithm, we stage a competition between $p$- and various anisotropic $h$-refinements, by computing the so-called guaranteed error reduction rate. The comparison between the various candidate refinements is based on the {\em error reduction rate $(e_{hp})$} defined as:
\be
e_{hp}: = \frac{ \Vert u - P_\mathrm{old} u \Vert^2 - \Vert u - P_\mathrm{new} u \Vert^2}{N_\mathrm{new} - N_\mathrm{old}}\, ,
\label{eq:rate}
\ee
where $u$ represents the {\em reference solution} obtained with the $hp$-refined mesh generated using the DPG residual, $P_\mathrm{old}$ is the $L^2$-projection onto the original coarse mesh element (space), $P_\mathrm{new}$ is the $L^2$-projection onto a {\em refined element} (space), $N_\mathrm{new}$ and $N_\mathrm{old}$
are the dimensions of the new and old spaces (number of dofs), respectively, and $\Vert \cdot \Vert$ denotes the $L^2$-norm over the considered element $K$.

\begin{figure}[!htb]
\centering
\begin{subfigure}[b]{0.2\textwidth}
	\includegraphics[width=\textwidth]{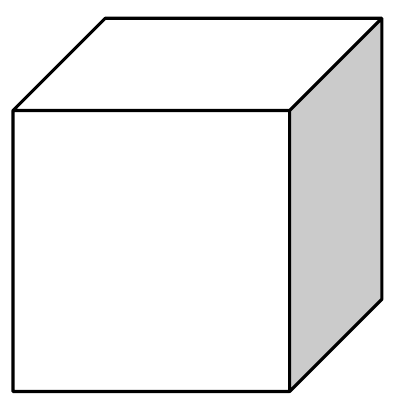}
	\caption{}
\end{subfigure} \hspace{0.02\textwidth}
\begin{subfigure}[b]{0.2\textwidth}
	\includegraphics[width=\textwidth]{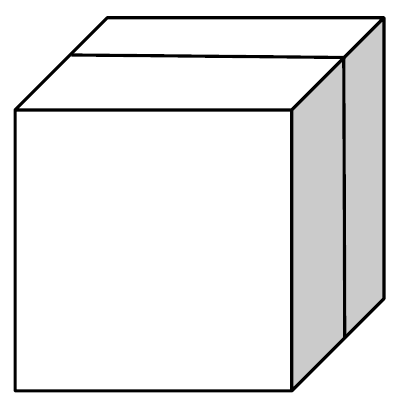}
	\caption{}
\end{subfigure} \hspace{0.02\textwidth}
\begin{subfigure}[b]{0.2\textwidth}
	\includegraphics[width=\textwidth]{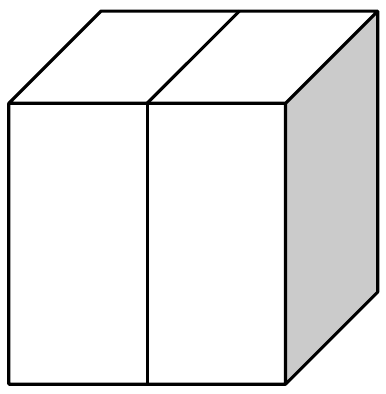}
	\caption{}
\end{subfigure} \hspace{0.02\textwidth}
\begin{subfigure}[b]{0.2\textwidth}
	\includegraphics[width=\textwidth]{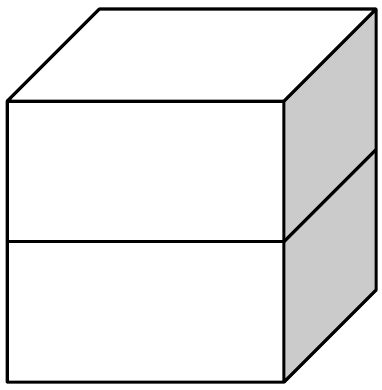}
	\caption{}
\end{subfigure}

\begin{subfigure}[b]{0.2\textwidth}
	\includegraphics[width=\textwidth]{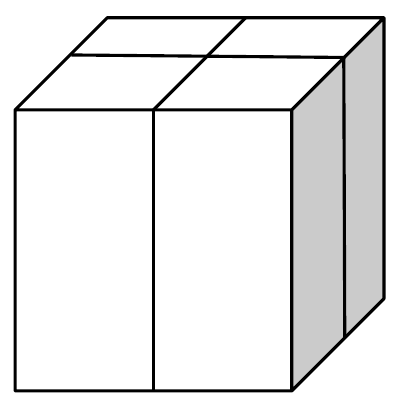}
	\caption{}
\end{subfigure} \hspace{0.02\textwidth}
\begin{subfigure}[b]{0.2\textwidth}
	\includegraphics[width=\textwidth]{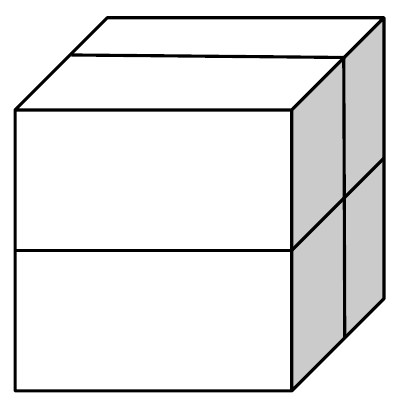}
	\caption{}
\end{subfigure} \hspace{0.02\textwidth}
\begin{subfigure}[b]{0.2\textwidth}
	\includegraphics[width=\textwidth]{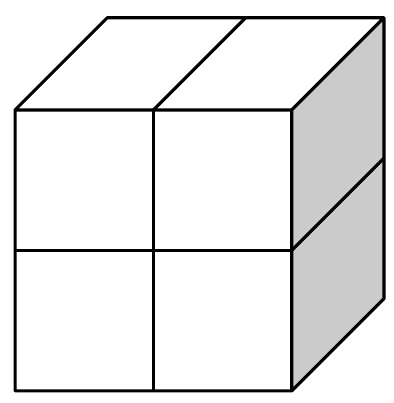}
	\caption{}
\end{subfigure} \hspace{0.02\textwidth}
\begin{subfigure}[b]{0.2\textwidth}
	\includegraphics[width=\textwidth]{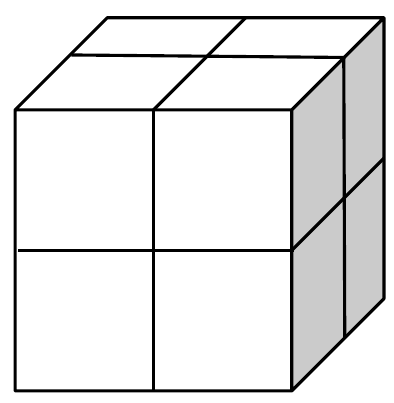}
	\caption{}
\end{subfigure}
	\caption{Various possible $h$-refinements for a hexahedral element, depicted in (a): (b--d) anisotropic $h_2$-refinements; (e--g) anisotropic $h_4$-refinements; and (h) isotropic $h_8$-refinement.}
	\label{anisohref}
\end{figure}

The optimal element refinement is determined by staging a competition among various candidate refinements. For hexahedral elements considered in this paper, there are eight possibilities: no $h$-refinement (i.e.~$p$-refinement only), three anisotropic $h_2$-refinements, three anisotropic $h_4$-refinements, and the isotropic $h_8$-refinement. \Cref{anisohref} illustrates all possible $h$-refinement candidates. Each of the eight refinements is accompanied with the determination of the optimal distribution of polynomial degrees. This leads to a catastrophically large number of possible cases for $hp$-refinement. With $p_x,p_y,p_z \in \{ 1,\ldots,10 \}$, there are ``only'' $10^3$ scenarios for the just $p$-refined element, but a staggering total of $10^{24}$ cases for the $h_8$-refined element. Clearly, a simple search through all possible cases is not feasible. 
Instead we rely on the classical $p$-refinement strategy, see e.g.~\cite{DOS84}, based on increasing the polynomial order in the subelement with the maximum error. This reduces the discrete search to the so-called {\em maximum error reduction path} through the vast discrete space of potentially possible refinements.

\begin{figure}[!htb]
\centering
\includegraphics[scale=0.5]{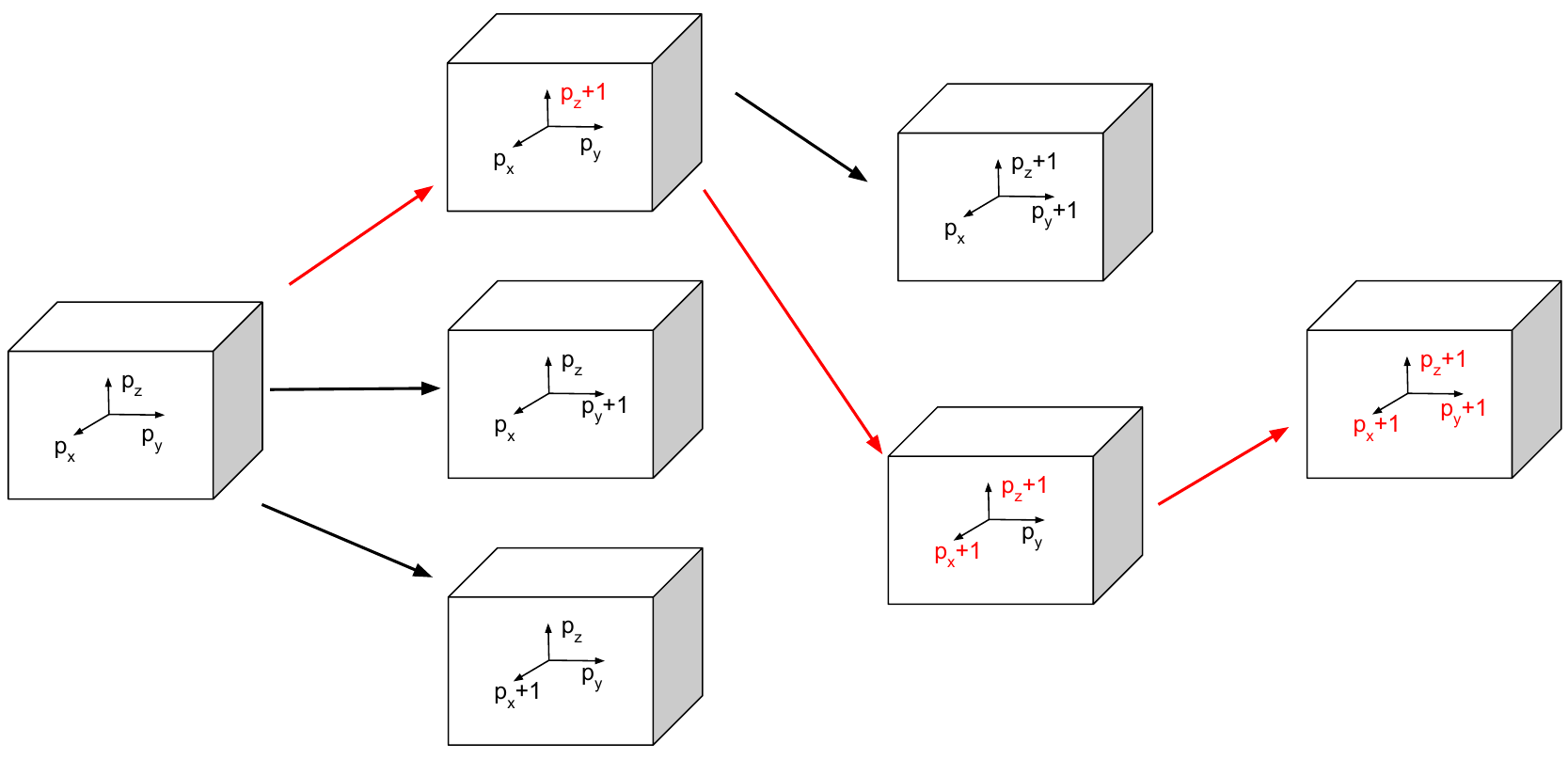}
\caption{Maximum error reduction path for the $p$-refined element: traversing from $(p_x,p_y,p_z)$ to $(p_x+1,p_y+1,p_z+1)$ for a hexahedral element.} \label{mep}
\end{figure}

\paragraph{Maximum error reduction path for a \texorpdfstring{$p$}{p}-refined element}
We begin the discussion with the simplest case: $p$-refinement only. Assuming that the polynomial order can only increase (by one order), there are only a total of $2^3 -1 = 7$ possible scenarios. The direct search is then possible but can be replaced with a slightly faster dynamic search, as illustrated in~\Cref{mep}. To choose the optimal $p$-refinement,  we traverse from $(p_x,p_y,p_z)$ to $(p_x+1,p_y+1,p_z+1)$ by increasing the order in directions that maximize $e_{hp}$. For a hexahedral element, the path of traversal has two stages. The first stage has three branches corresponding to $p_x,p_y$, and $p_z$. The second stage has two branches corresponding to the remaining directions,  with the final configuration being $(p_x+1,p_y+1,p_z+1)$. In~\Cref{mep}, the arrows in red represent the branches corresponding to the highest values of $e_{hp}$ at each stage, and the polynomial order marked in red indicates the polynomial order increased after each stage.  

Following the path, we select the $p$-refinement that delivers the largest error reduction rate. In the case of an affine element, the element Jacobian $(\mathrm{jac})$ is constant, and the $L^2$-Piola transform (pullback map) reduces to a scaling with the Jacobian:
\be
\phi_j(x) = \frac{1}{\mathrm{jac}} \hat{\phi}_j(\xi) , \qquad \mathrm{jac} = \left\vert \frac{\ptl x_i}{\ptl \xi_j} \right\vert,
\ee
where $\phi_j$ is an element $L^2$ shape function corresponding to a master element shape function $\hat{\phi}_j$.
Consequently,  the $L^2$ mass matrix,
\be
M_{ij} := \int_K \phi_i \phi_j \, dx = \frac{1}{ \mathrm{jac}} \int_{\hat{K}} \hat{\phi}_i \hat{\phi}_j \, d \xi \, ,
\ee
is diagonal, and the  evaluation of the $L^2$ projection of a function $u$ onto  a subspace spanned by functions $\phi_1,\ldots,\phi_N$, reduces to the evaluation
of the load vector:
\be
P_N u = \sum_{j=1}^N u_j \phi_j,\qquad u_j = 
\frac{1}{M_{jj}} \int_K u \phi_j \, dx \, .
\ee
Raising the polynomial order in one direction amounts to adding extra orthogonal shape functions $\phi_{N+l}$ with $l = 1,....,n$. Consequently, evaluation of the error reduction rate reduces to:
\begin{align}
\frac{\Vert u - P_N u \Vert^2 - \Vert u - P_{N+1} u \Vert^2}{n} 
& = \frac{\Vert P_{N+1} u \Vert^2 - \Vert P_N u \Vert^2}{n} \\
& = \frac{1}{n} \sum_{l=1}^{n}\vert u_{N+l} \vert^2 M_{N+l,N+l}
=  \frac{1}{n} {\sum_{l=1}^{n} \left( \frac{\int_K u \phi_{N+l} \, dx}{M_{N+l,N+l}}\right)^2 M_{N+l,N+l}} \nonumber \\[5pt]
& = \frac{1}{n} \sum_{l=1}^{n} M_{N+l,N+l}^{-1} \left( \int_K u \phi_{N+l} \, dx \right)^2 \, . \nonumber
\end{align}
In the case of a general curvilinear element, the $L^2$ mass matrix is not diagonal, and we use the telescopic solver based on the Cholesky decomposition described in \cite[p.~140]{LDbook_b}.

\paragraph{Maximum error reduction path for an \texorpdfstring{$h$}{h}-refined element}
Contrary to the pure $p$-refinement, we always start with a trilinear element where $p_x = p_y = p_z =1$. The reference solution $u$ is projected onto the subelement mesh and, based on the distribution of the error, subelements are selected for refinement using a {\em greedy strategy} with a $70\%$ factor. Once the subelements have been identified for $p$-refinement, the routine described above is employed to determine the optimal $p$-refinement for each subelement.

\begin{figure}[!htbp]
\centering
\begin{subfigure}[t]{0.33\textwidth}
	\includegraphics[width=\textwidth]{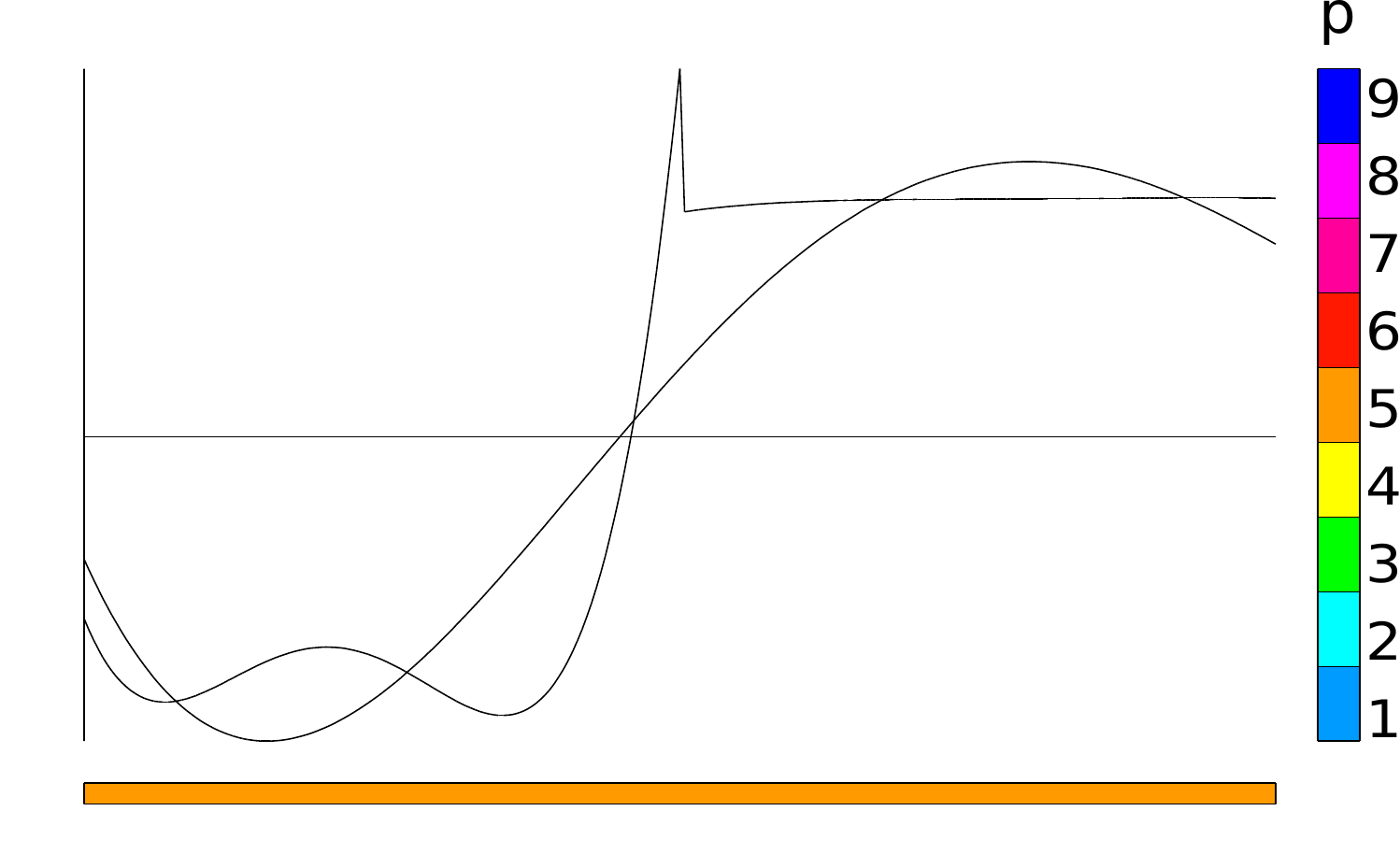}
	\caption{Purely $p$-refined element: \\
	$p_K = 5$ \\
	$e_{hp} = 0.01952$}
	\label{1Dpref}
\end{subfigure}
\begin{subfigure}[t]{0.33\textwidth}
	\includegraphics[width=\textwidth]{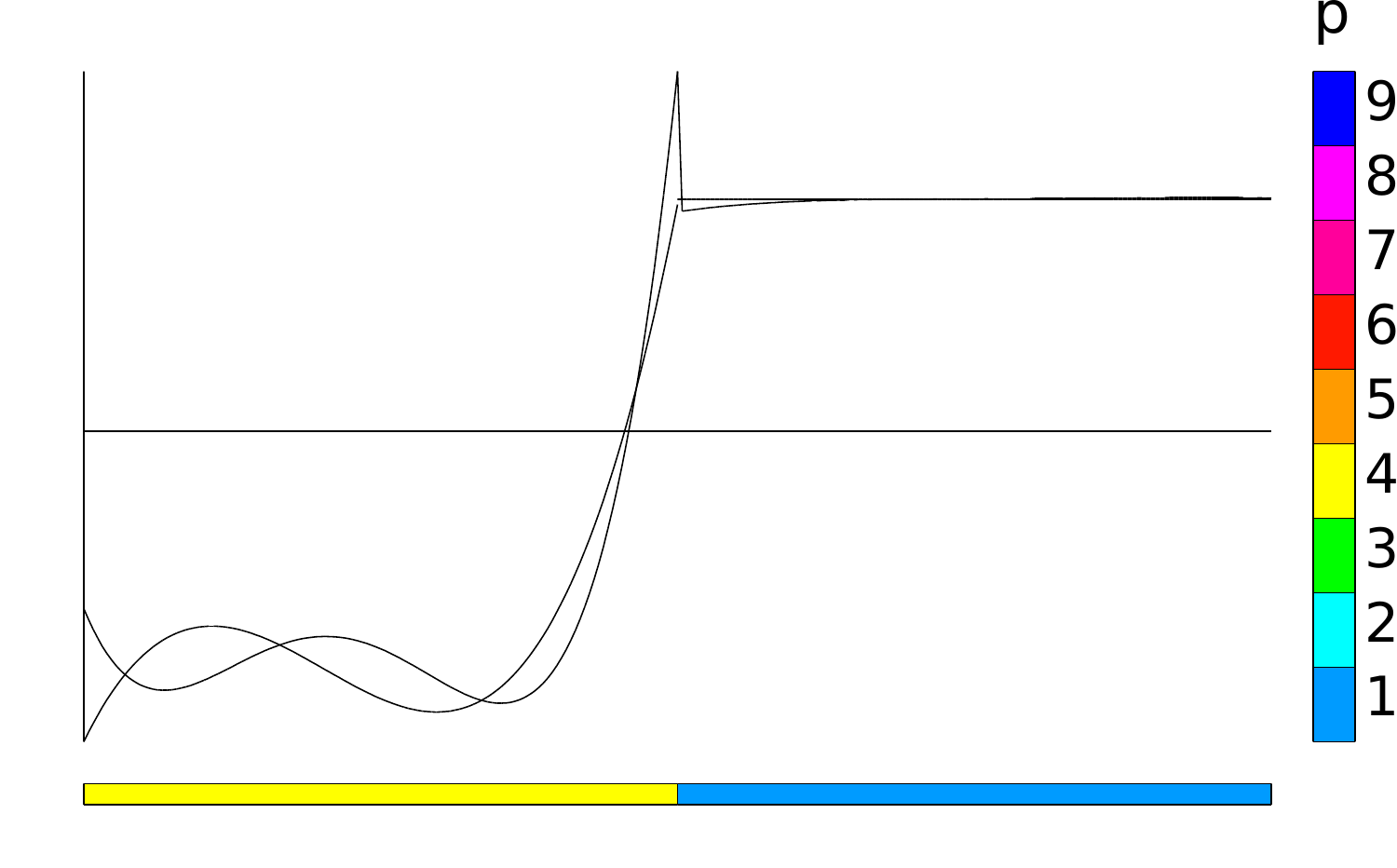}
	\caption{Winning $h$-refined element:\\ 
	$(p_{K_1}, p_{K_2}) = (4,1)$\\
	$e_{hp} = 0.23196$}
	\label{1Dhref}
\end{subfigure}
\begin{subfigure}[t]{0.32\textwidth}
	\centering
	\includegraphics[width=0.9\textwidth,trim={60pt 0pt 80pt 0pt},clip]
	{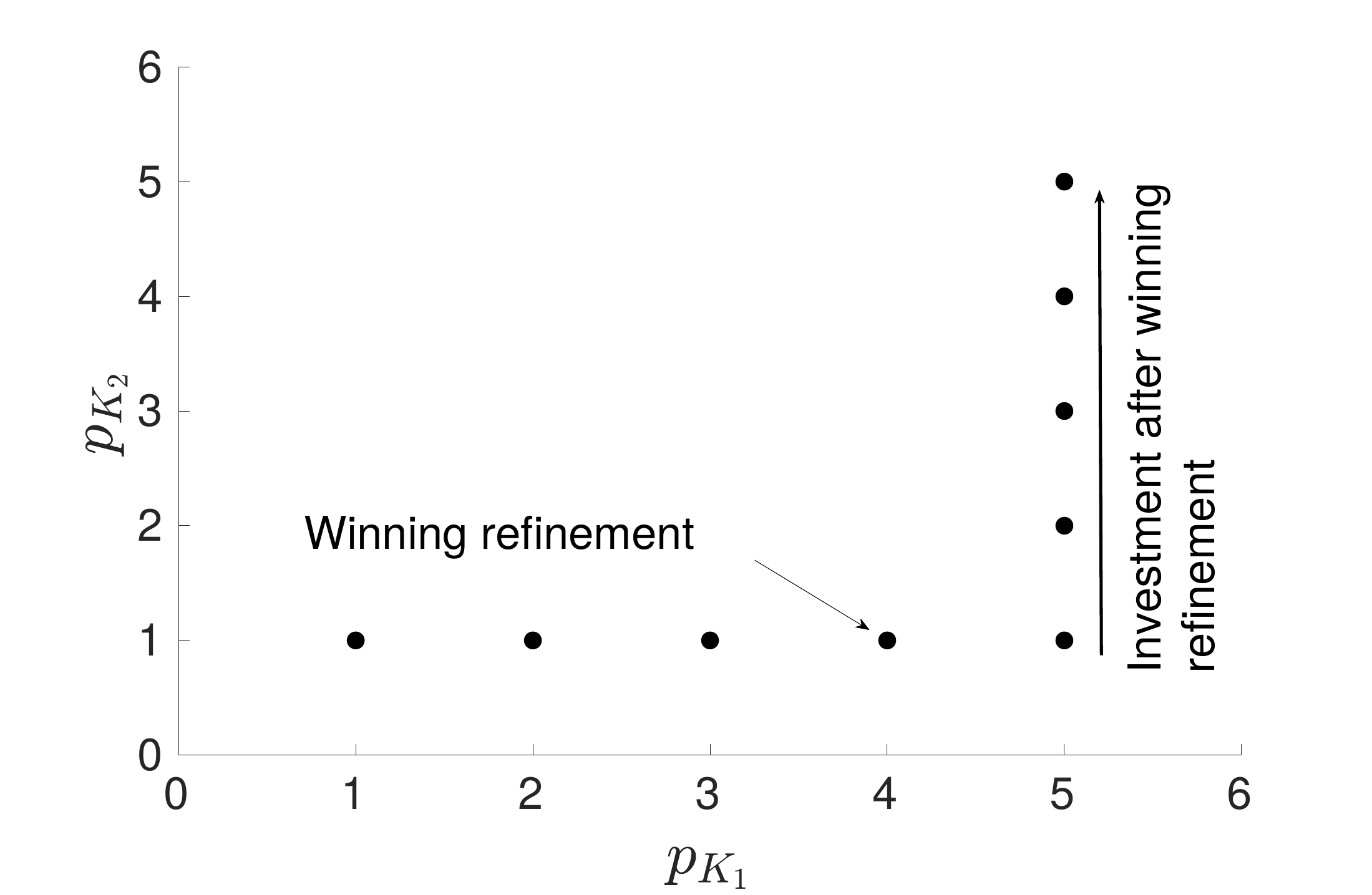}
	\caption{Refinement path of the\\ $h$-refined element}
	\label{1Dpath}
\end{subfigure}
\caption{Staging a competition between the $p$-refined and $h$-refined element. The maximum error reduction path for the $h$-refined element traverses from $(p_{K_1}, p_{K_2}) = (1,1)$ to the winning refinement $(p_{K_1}, p_{K_2}) = (4,1)$.}
\label{1Dexample}
\end{figure}

\Cref{1Dexample} shows the simple case of a 1D element $K$, starting with polynomial order $p_K=4$. The subelements of the $h$-refined element $K$ are denoted $K_1$ and $K_2$, and their respective polynomial orders $p_{K_1}$ and $p_{K_2}$. The maximum error reduction path for this case (illustrated in \Cref{1Dpath}) leads to the winning refinement $(p_{K_1},p_{K_2}) = (4,1)$ with the approximate solution shown in \Cref{1Dhref}.

\paragraph{The optimal refinement} The selection of the optimal refinement is carried out by comparing the best error reduction rates delivered by the eight differently $h$-refined meshes. The highest error reduction rate, delivered by the optimal refinement, is called the {\em guaranteed error reduction rate} and denoted by $e_{hp}^*$.

\subsection{Step 2: Determining Which Elements to Refine}
We loop over all considered coarse mesh elements to determine the element with the {\em best guaranteed error reduction rate} $e_{hp,\text{max}}^*$. In principle, one could then refine only this one element. However, to accelerate the refinements (i.e.~reduce the number of refinement steps), a greedy strategy is employed selecting all elements that deliver a rate greater than or equal to $25\%$ of the best guaranteed error reduction rate. Note that this strategy implies that there may be elements initially marked for refinement by the DPG residual which ultimately remain unrefined.

\subsection{Step 3: Determining the Final Refinements}
For each element selected for refinement in Step~2, we could simply execute the corresponding optimal refinement determined in Step~1; and we do this indeed for the purely $p$-refined elements. However, when performing $h$-refinements we typically choose to invest additional dofs by considering the already-performed $p$-refinements that followed the optimal refinement while investigating error reduction rates in Step~1.

In particular, in Step~1 we recorded the error reduction rates for all subelement meshes following the maximum error reduction path. On this path, we select the maximum investment (in terms of new dofs) that still delivers $25\%$ of the best guaranteed error reduction rate (meaning it would still satisfy the Step~2 criterion). The rational for doing so is to reduce the overall number of outer-loop iterations (number of refinement steps) by maximizing the investment in each step as long as it pays off sufficiently (delivering a sufficiently high error reduction rate, as determined by Step~2).

For example, in the 1D case illustrated in~\Cref{1Dexample}, the refinement shown in~\Cref{1Dhref} won the competition with the $p$-refinement (\Cref{1Dpref}) but, dependent upon the threshold value used in the greedy strategy, we may choose to invest additional dofs in one of the subelements.

Next, we consolidate Steps~1--3 and present the mesh optimization algorithm. In~\Cref{algo}, $\text{tol}$ denotes the user-provided tolerance value for the DPG residual.

\begin{algorithm}[htb]
\caption{Mesh Optimization Algorithm} \label{algo}
\begin{algorithmic}[1]

\State Start with an initial trial mesh
\While{$\eta_{\Omega_h} > \text{tol}$}
\State Solve the problem on the current mesh.
\State Compute the DPG residual for the current mesh: $\eta_{\Omega_h} = {\left(\sum_{K \in \Omega_h} \eta_K \right)}^{1/2}$.
\State Use the element residuals $(\eta_k)$ to mark elements for refinements (D\"{o}rfler strategy).

\State Isotropically $hp$-refine marked elements to generate the fine mesh.
\State Compute the reference solution $u$ using the fine mesh.
\State {\em Step 1:} For each refined element $K$: \\
	\hspace{1cm} Determine the best possible $p$-refinement using the maximum error reduction path.\\
	\hspace{1cm} Determine the best possible $h$-refinement using the maximum error  reduction path. \\
	\hspace{1cm} Use error reduction rates to decide between $p$- and $h$-refinement.\\
	\hspace{1cm} Determine the element guaranteed error reduction rate $(e^*_{hp,K})$.

	\State {\em Step 2:} Determine the best guaranteed error reduction rate $(e^*_{hp,\text{max}})$.
	\State Unrefine the mesh.

	\State {\em Step 3:} For each element $K$ marked for refinement:
	\If{$e^*_{hp,K} \geq 0.25 \, e^*_{hp,\text{max}}$}
		\State Perform the optimal $hp$-refinement.
	\EndIf
\EndWhile
\end{algorithmic}
\end{algorithm}

%
%

\subsection{Mesh Closure}

The $hp$ algorithm is implemented in $hp$3D, a general-purpose finite element code supporting hybrid meshes consisting of elements of all shapes (hexas, tets, prisms, pyramids), conforming discretizations of the exact-sequence spaces ($H^1$-, $\Hcurl$-, $\Hdiv$-, and $L^2$-conforming elements), solution of coupled multiphysics problems, and {\em anisotropic} $hp$-refinements \cite{Henneking_Demkowicz_book, henneking2022hp3d}. $hp$3D supports MPI/OpenMP parallelism \cite{henneking2021phd} and is available under BSD-3 license.\footnote{\url{https://github.com/Oden-EAG/hp3d}} In the code, any $h$-refinement is executed in two steps. Given a list of elements to refine (along with the requested, possibly anisotropic, $h$-refinement flags), we proceed as follows.
\begin{description}
\item[Closure step 1 (local):] Refine the elements from the list in the provided order, enforcing two rules:
\begin{itemize}
\item{\bf Compatibility with existing face refinements:} upgrade the requested element refinement flag to accommodate {\em existing face refinements}.
\item{\bf One-irregularity rule for faces:} employ the standard {\em shelf} or {\em queue} algorithm (\cite[p.~71]{LD_b}) to ensure that no face is refined unless the face\footnote{More precisely, the mid-face node.} is unconstrained.
\end{itemize}
If one of the element faces is constrained, the element is placed on the shelf, and a necessary refinement of the neighbor across the face is executed, to eliminate the constraint. If the one-irregularity rule for faces prohibits the refinement, the corresponding neighbor is placed on the shelf and so on. Once the refinement of the processed element is possible, it is executed and the process resumes with the last element from the shelf. The algorithm proceeds until the shelf is empty. All mesh manipulations (refinements) are supported for meshes that satisfy the one-irregularity rule for faces (not necessary for edges and vertices).
\item[Closure step 2 (global):] Loop through all elements and perform additional necessary refinements to eliminate edges and vertices with multiple constraints.
\end{description}
We refer to \cite{Henneking_Demkowicz_book} {for} a more formal exposition of the algorithms. In the end, in both steps, a number of additional, {\em unwanted} refinements may be executed. These refinements can be {\em isotropic} or {\em anisotropic}, reflecting minimal requirements to eliminate the nodes with multiple constraints. In the `global' $hp$-refinement driven by the DPG residual, all unwanted refinements are chosen to be isotropic. This is motivated by the fact that an unwillingly refined element (in Step 1) may, in fact,  be on the DPG list of wanted refinements. However, once the optimal $hp$-refinements are determined, all unwanted refinements are executed in a minimal, anisotropic way.

All unwillingly $h$-refined elements retain their respective polynomial order. In principle, one could attempt to find the corresponding optimal distribution of polynomial orders, but this has been not done in our current implementation. Hence, the presented meshes may be slightly overrefined.

%
%
\section{Numerical Results
\label{sec:results}
}
\subsection{A Boundary Layer Problem}
Sharp boundary layers are among the most commonly encountered flow features in computational fluid dynamics. Our first numerical experiment demonstrates the proposed algorithm's efficacy in resolving such boundary layers. In this test case, we solve a Poisson problem with a manufactured solution containing boundary layers. The manufactured solution is a three-dimensional extension of the solution of the Egger-Sch\"{o}berl problem \cite{esch}. In particular, we solve
\begin{align}
\begin{split}
-\nabla^2  u &= f(x,y,z) \quad \text{in} \quad \Omega:={(0,1)}^3, \\
u &= 0 \, \, \qquad \qquad \text{on} \quad \Gamma_u, \\
\nabla u \cdot \bm{n} &= g(x,y,z) \quad \text{on} \quad  \Gamma_{\sigma},
\end{split} \label{bl}
\end{align}
where
\be
\begin{split}
\Gamma_u &=  \left( [0,1) \times [0,1) \times \{0\} \right) \cup \left( [0,1) \times \{0\} \times [0,1) \right) \cup \left( \{0\} \times [0,1) \times [0,1) \right) , \\
\Gamma_{\sigma} &= \left( [0,1] \times [0,1] \times \{1\} \right) \cup \left( [0,1] \times \{1\} \times [0,1] \right) \cup \left( \{1\} \times [0,1] \times [0,1] \right).
\end{split}
\ee

In~(\ref{bl}), $\bm{n}$ is the outward normal, and $f$ and $g$ are generated using the exact solution. The exact solution is given by
\be
u(x,y,z) = \left( x + \frac{e^{x/\epsilon} - 1}{1 - e^{1/\epsilon}} \right)\left( y + \frac{e^{y/\epsilon} - 1}{1 - e^{1/\epsilon}} \right)\left( z + \frac{e^{z/\epsilon} - 1}{1 - e^{1/\epsilon}} \right).
\ee

The solution exhibits a boundary layer near $x \approx 1$, $y \approx 1$ and $z \approx 1$. The strength of the boundary layer is inversely proportional to $\epsilon$. In this numerical experiment, $\epsilon = 0.005$. The $hp$-adaptation is initialized with a mesh comprising only eight elements with a constant polynomial order of $(2,2,2)$.\footnote{In $hp$3D, we employ exact-sequence spaces \cite{lec_LD_FEM}. Hence, an order of $(p_x,p_y,p_z)$ denotes $L^2$ shape functions of order $(p_x-1,p_y-1,p_z-1)$.} 

\begin{figure}[!htb]
\centering
\begin{subfigure}{0.48\textwidth}
	\centering
	\includegraphics[width=\textwidth, trim={0pt 80pt 0pt 80pt}, clip]
	{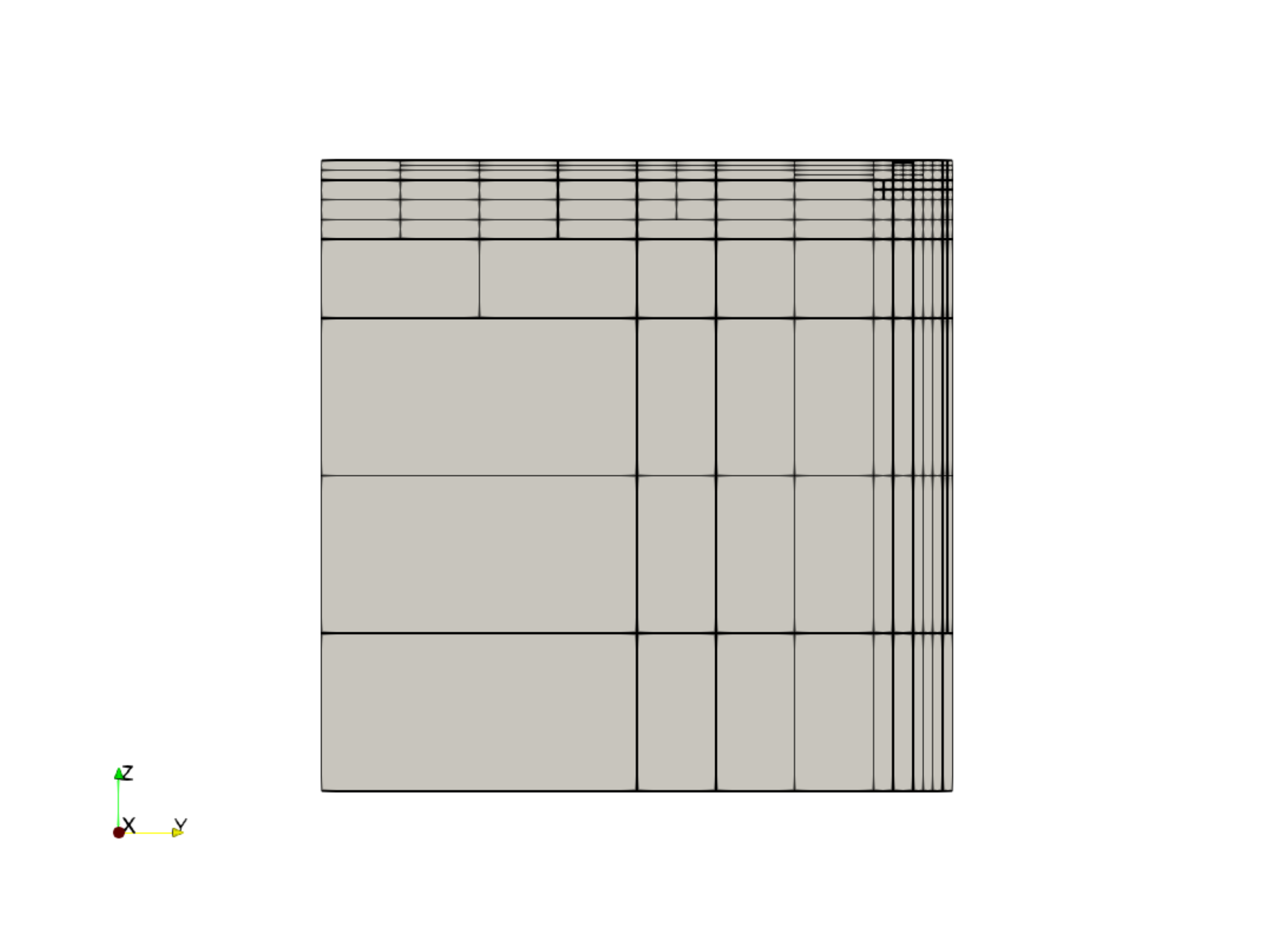}
	\caption{Cross-section of the mesh at $x = 0.95$} \label{mesh_es}
\end{subfigure}
\begin{subfigure}{0.48\textwidth}
	\centering
	\includegraphics[width=\textwidth, trim={0pt 80pt 0pt 80pt}, clip]
	{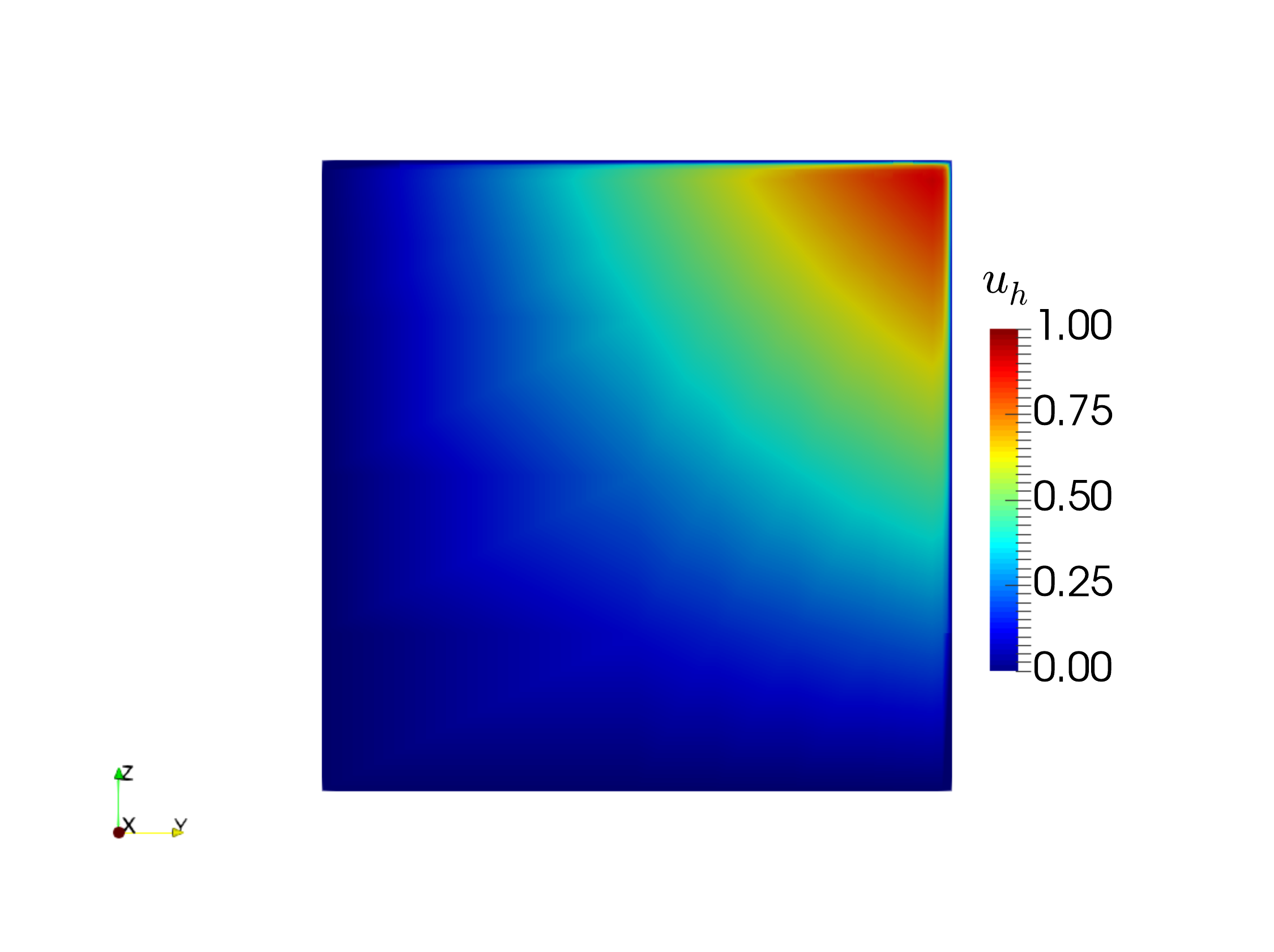}
	\caption{Contour plot of the solution at $x = 0.95$} \label{sol_es}
\end{subfigure}
\caption{Boundary layer problem: (a) cross-section of the mesh showing anisotropic elements required to resolve the boundary layers; and (b) contour plot illustrating the boundary layers on the $yz$-plane. The boundary layers are along right and top faces of the cross-section.}
\end{figure}

\begin{figure}[!htb]
\centering
\begin{subfigure}{0.32\textwidth}
	\centering
	\includegraphics[width=\textwidth]{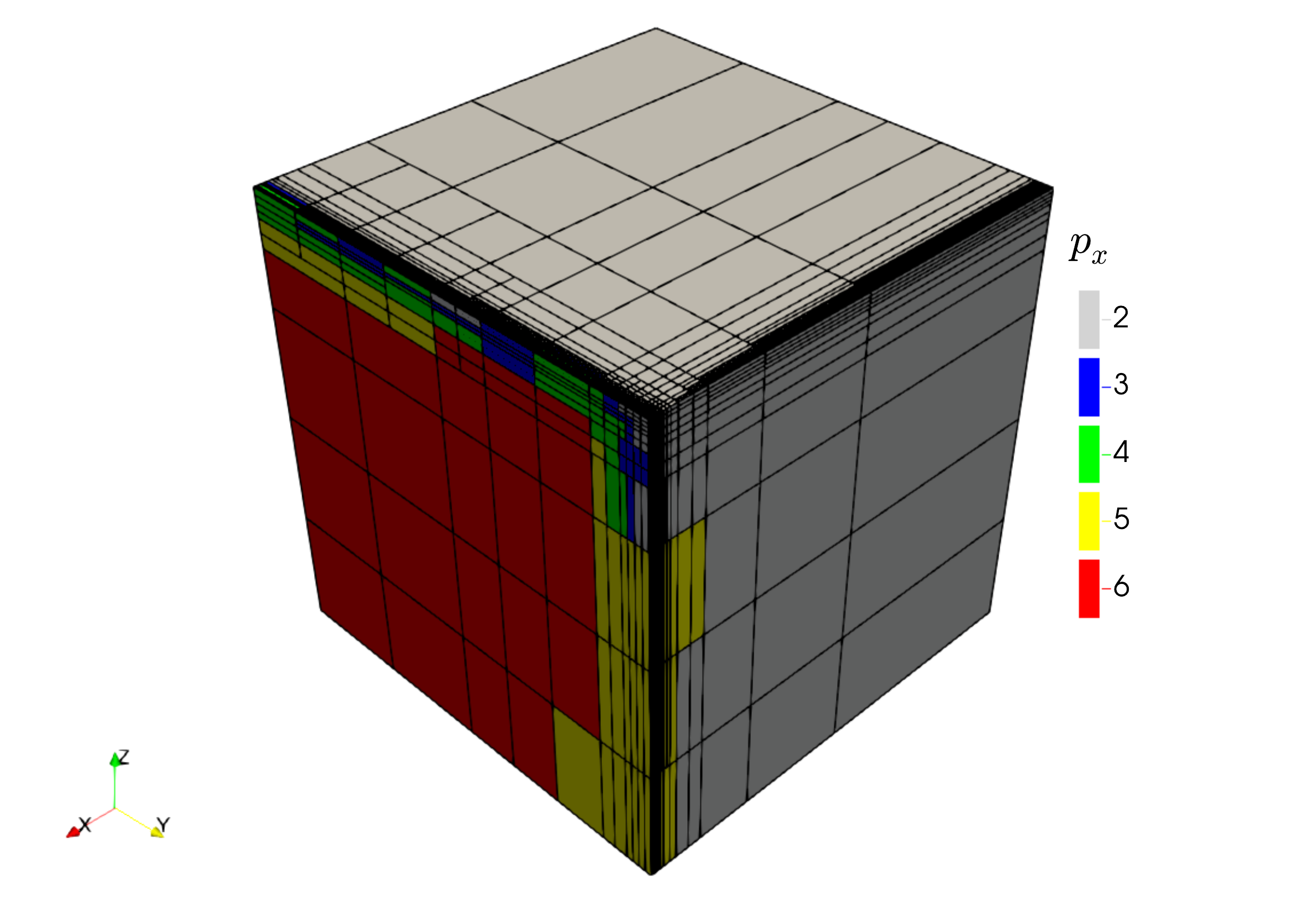}
	\caption{Polynomial order $p_x$}
\end{subfigure}
\begin{subfigure}{0.32\textwidth}
	\centering
	\includegraphics[width=\textwidth]{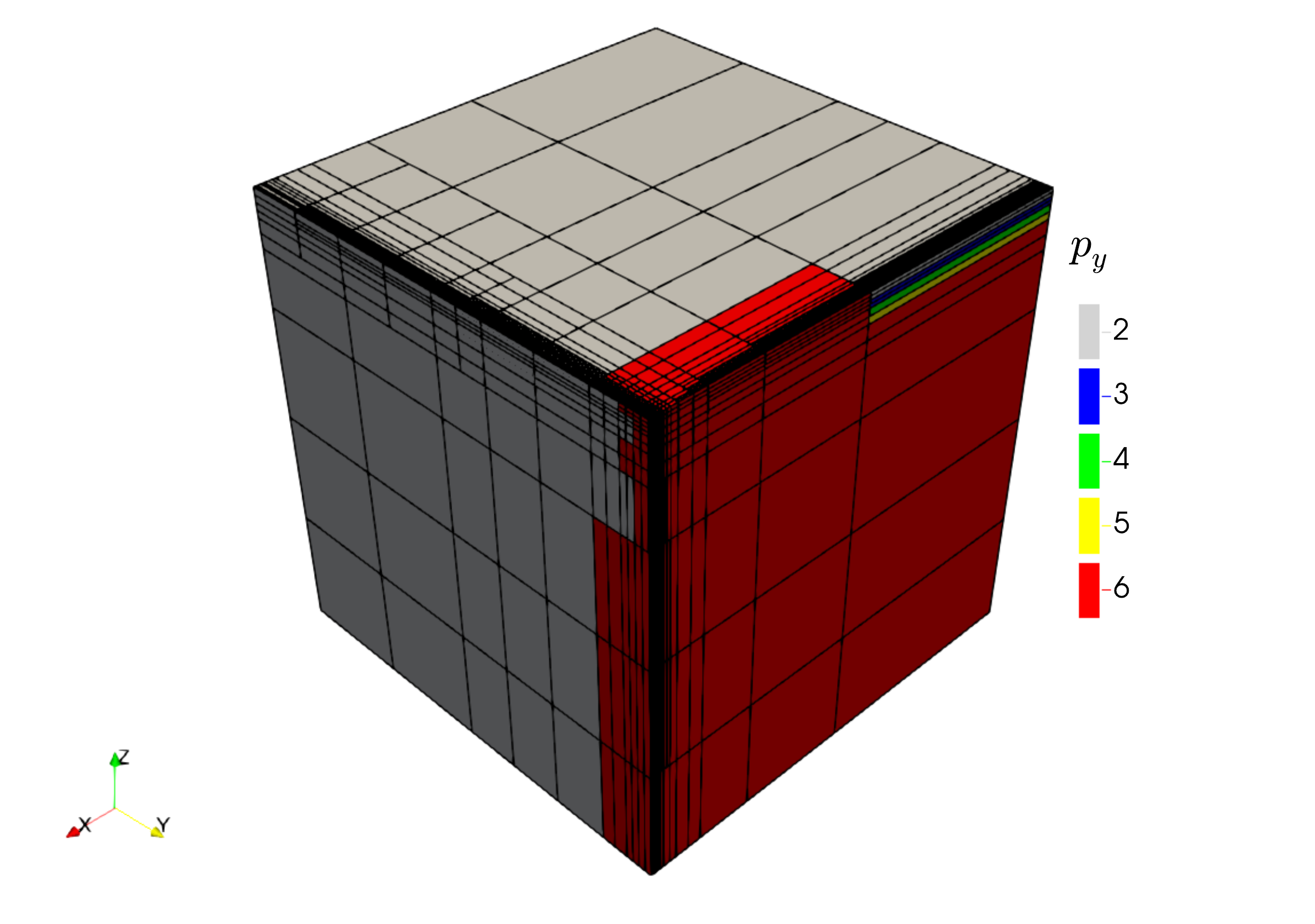}
\caption{Polynomial order $p_y$}
\end{subfigure}
\begin{subfigure}{0.32\textwidth}
	\centering
	\includegraphics[width=\textwidth]{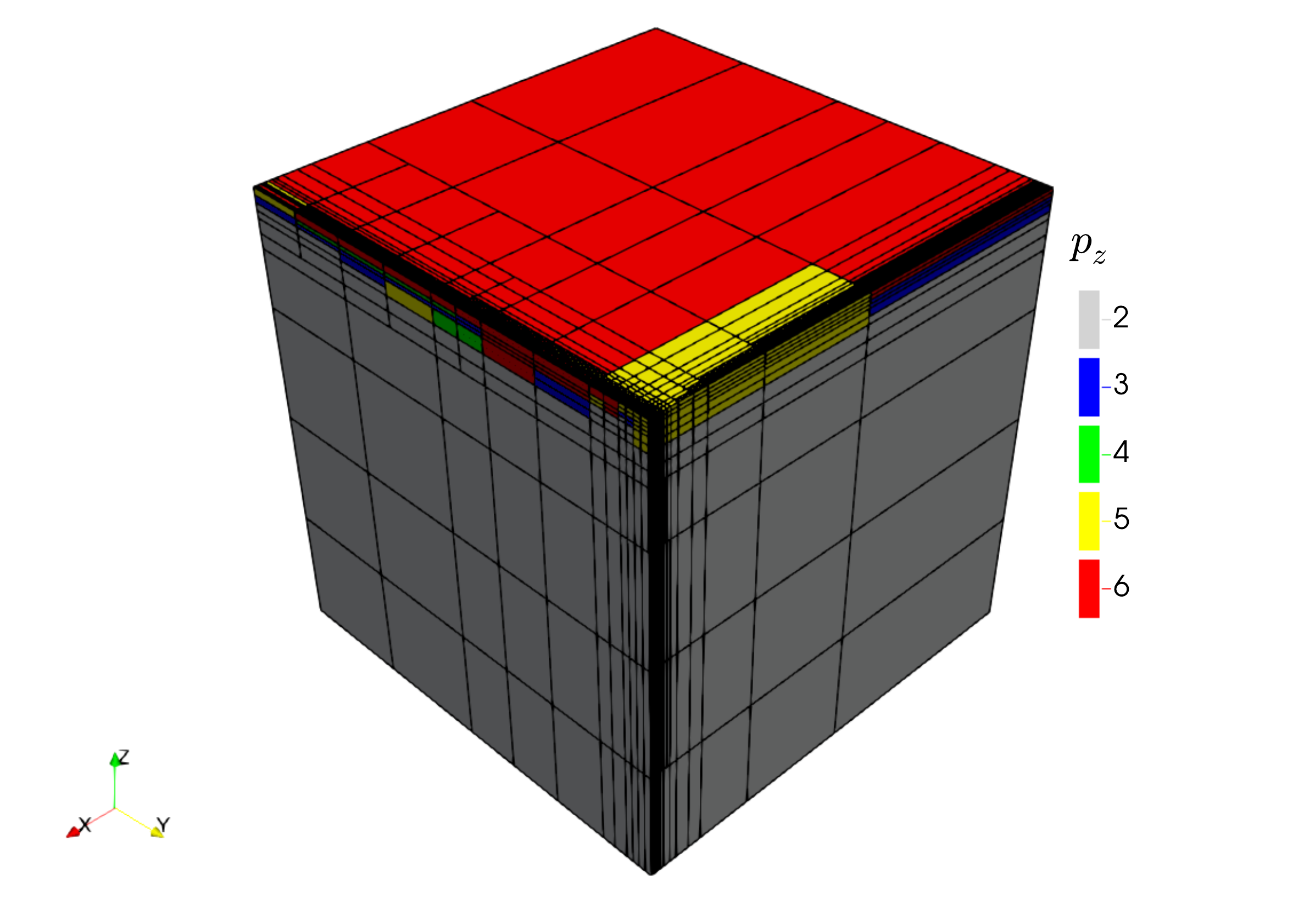}
	\caption{Polynomial order $p_z$}
\end{subfigure}
\caption{Boundary layer problem: an adapted mesh with $855\,532$ dofs; coloring indicates the polynomial distributions $p_x$, $p_y$, $p_z$ in $x$-, $y$-, $z$-direction, respectively. The algorithm prescribes higher-order polynomials anisotropically corresponding to each boundary layer along the $x$-, $y$-, and $z$-axis.} \label{orth_mesh}
\end{figure}

\begin{figure}[!htb]
\centering
\resizebox{0.45\textwidth}{!}{
\begin{tikzpicture}
		\begin{semilogyaxis}[xmin=6,xmax=210, ymin=1e-4,ymax=200,xlabel=\large{$\sqrt[3]{\text{ndof}}$},ylabel=\large{$\text{Relative } L^2 \text{ error}$},grid=major,legend style={at={(1,1)},anchor=north east,font=\small,rounded corners=2pt}]
		\addplot[color = blue,mark=square*] table[x= ndof_tot, y=err_L2_rel, col sep = comma] {results/ES_testcase_invst/error_p2_ES_results.txt};
		\addplot [color = red,mark=square*] table[x= ndof_tot, y=err_L2_rel, col sep = comma] {results/ES_testcase_invst/error_p3_ES_results.txt};
		\addplot [color = black,mark=square*] table[x= ndof_tot, y=err_L2_rel, col sep = comma] {results/ES_testcase_invst/error_p4_ES_results.txt};
		\addplot [color = magenta,mark=square*] table[x= ndof_tot, y=err_L2_rel, col sep = comma] {results/ES_testcase_invst/error_hp_ES_invst_results.txt};
		\legend{$p = 2$,$p = 3$,$p = 4$,$hp$}
		\end{semilogyaxis}
	\end{tikzpicture}
}
\hspace{0.05\textwidth}
\resizebox{0.45\textwidth}{!}{
\begin{tikzpicture}
		\begin{semilogyaxis}[xmin=6,xmax=210, ymin=1e-4,ymax=200,xlabel=\large{$\sqrt[3]{\text{ndof}}$},ylabel=\large{$\text{Residual}$},grid=major,legend style={at={(1,1)},anchor=north east,font=\small,rounded corners=2pt} ]
		\addplot[color = blue,mark=square*] table[x= ndof_tot, y=res, col sep = comma] {results/ES_testcase_invst/error_p2_ES_results.txt};
		\addplot [color = red,mark=square*] table[x= ndof_tot, y=res, col sep = comma] {results/ES_testcase_invst/error_p3_ES_results.txt};
		\addplot [color = black,mark=square*] table[x= ndof_tot, y=res, col sep = comma] {results/ES_testcase_invst/error_p4_ES_results.txt};
		\addplot [color = magenta,mark=square*] table[x= ndof_tot,  y=res, col sep = comma] {results/ES_testcase_invst/error_hp_ES_invst_results.txt};
		\legend{$p = 2$,$p = 3$,$p = 4$,$hp$}
		\end{semilogyaxis}
\end{tikzpicture}
}
\caption{Boundary layer problem: convergence of relative $L^2$ error and DPG residual. Even though there is a marginal decrease in the rate of convergence for the $hp$-refinements, both the error and the residual are $2$--$3$ orders of magnitude lower compared to the $h$-refinements for approximately the same number of dofs.} \label{errConvegebl}
\end{figure}
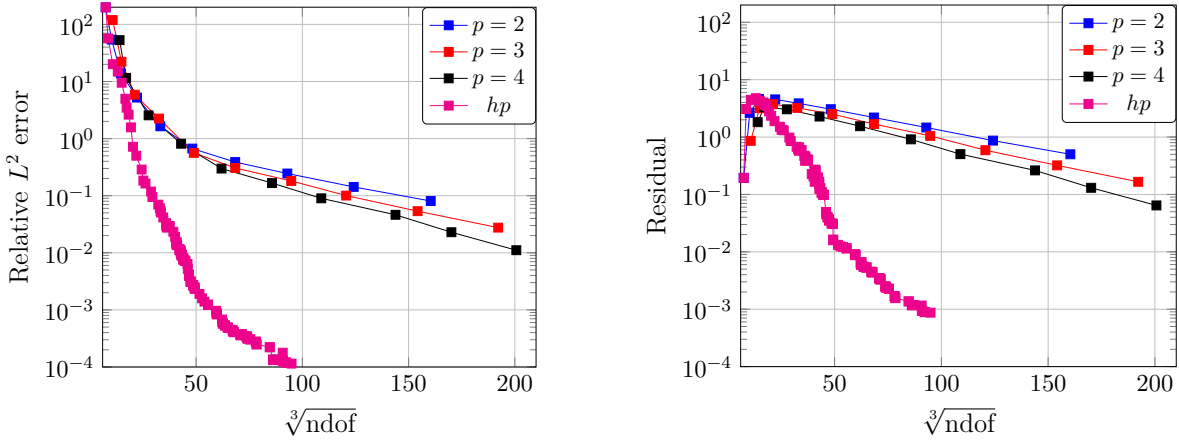

\Cref{mesh_es,sol_es} display the cross-section of an adapted mesh and the corresponding solution contour, respectively. \Cref{orth_mesh} depicts the polynomial distribution around the boundary layers on an anisotropically adapted $hp$-mesh. \Cref{errConvegebl} presents the convergence results, comparing isotropic $h$-adaptation and the proposed $hp$-refinement strategy. The D\"{o}rfler parameter for both isotropic and $hp$-refinement is $0.75$. In~\Cref{errConvegebl}, the depicted error is the combined relative error in all $L^2$ variables.

\Cref{orth_mesh} clearly illustrates the strong anisotropy and grading in the element size and the polynomial distribution. The anisotropy and the grading in element size are paramount for resolving strong boundary layers efficiently. The algorithm also prescribes an anisotropic polynomial distribution in the boundary layers instead of an isotropic one. This directional preference of prescribing polynomial orders showcases a significant advantage of the proposed $hp$-refinement strategy: the ability to complement an anisotropic $h$-refinement with an anisotropic $p$-refinement. This approach makes the refinement strategy highly efficient in terms of allocating dofs when the solution exhibits strong anisotropic features. The algorithm does not waste any dofs in directions where the solution variables do not exhibit significant variations.

From~\Cref{errConvegebl}, it is evident that anisotropic $hp$-refinements outperform isotropic $h$-refinements by orders of magnitude. The convergence plots show the error and the residual against $\sqrt[3]{\text{ndof}}$ (where $\text{ndof}$ represents the number of degrees of freedom), verifying exponential convergence. In~\Cref{errConvegebl}, a reduction in the convergence rate for the $hp$-refinement can be observed. The slowdown in convergence occurs due to the limiting of the highest polynomial order in the numerical experiments to  $p = 6$. 
The adaptation cycles are initially dominated by $h$-refinements. Once the boundary layers are resolved, the algorithm starts preferring both $p$-refinements along with $h$-refinements. This behavior is expected, since, increasing the polynomial order on coarse meshes while approximating solutions with high gradients can induce spurious oscillations.

\subsection{Fichera Cube Problem}
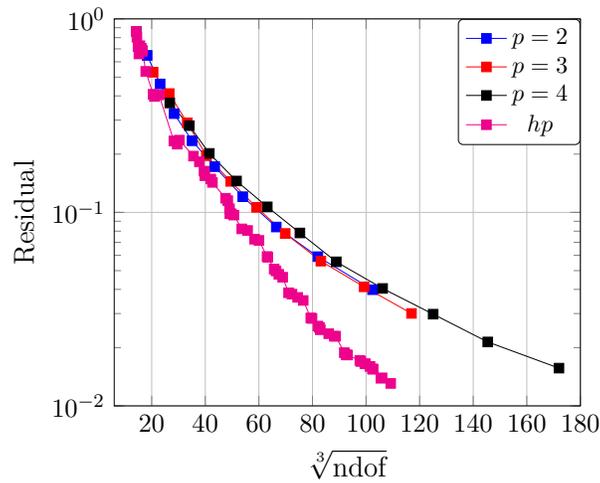
\begin{figure}[!htb]
\centering
\resizebox{0.5\textwidth}{!}{
\begin{tikzpicture}
		\begin{semilogyaxis}[xmin=6,xmax=180, ymin=1e-2,ymax=1.0,xlabel=\large{$\sqrt[3]{\text{ndof}}$},ylabel=\large{$\text{Residual}$},grid=major,legend style={at={(1,1)},anchor=north east,font=\small,rounded corners=2pt} ]
		\addplot[color = blue,mark=square*] table[x= ndof_tot, y=res, col sep = comma] {results/FC_testcase_invst/error_p2_FC_results.txt};
		\addplot [color = red,mark=square*] table[x= ndof_tot, y=res, col sep = comma] {results/FC_testcase_invst/error_p3_FC_results.txt};
		\addplot [color = black,mark=square*] table[x= ndof_tot, y=res, col sep = comma] {results/FC_testcase_invst/error_p4_FC_results.txt};
		\addplot [color = magenta,mark=square*] table[x= ndof_tot,  y=res, col sep = comma] {results/FC_testcase_invst/error_hp_FC_invst_results.txt};
		\legend{$p = 2$,$p = 3$,$p = 4$,$hp$}
		\end{semilogyaxis}
\end{tikzpicture}
}
\caption{Fichera cube problem: convergence of the DPG residual.} \label{resConvegefc}
\end{figure}

To demonstrate the efficacy of the proposed refinement strategy in the presence of multiple singularities, we solve the well-known Fichera cube problem and perform $hp$-adaptations using the proposed refinement strategy. The variant of the Fichera cube problem being solved here is given by:
\begin{align}
\begin{split}
\nabla^2 u &= 0 \quad \quad \qquad \,\, \text{in} \quad \Omega:={(-1,1)}^3 \setminus {[0,1]}^3, \\
u &= 0 \, \, \qquad \qquad \text{on} \quad \Gamma_u,\\
\nabla u \cdot \bm{n} &= g(x,y,z) \quad \text{on} \quad \Gamma_{\sigma}.
\end{split}
\end{align}
The domain is created by subdividing a large cube ${(-1,1)}^3$ into eight smaller cubes and then removing one of the cubes. The Dirichlet data $u = 0$ is imposed on the three square faces aligned with planes of coordinate axes, i.e.
\be
\Gamma_u = \left( [0,1] \times [0,1] \times \{0\} \right) \cup \left([0,1] \times \{0\} \times [0,1]\right) \cup \left(\{0\} \times [0,1] \times [0,1]\right).
\ee
The volumetric load for the problem is zero. The problem is driven by the Neumann boundary condition on $\Gamma_{\sigma}$ composed of the remaining faces of the cube. The data $g$ correspond to the sum of two-dimensional exact solutions of the L-shaped domain problem on $xy$-, $yz$-, and $xz$-planes. The exact solution of the L-shaped domain problem is given by:
\begin{align}
u_{\eta,\xi} = r^{\frac{2}{3}}cos(\theta), \quad r = \sqrt{\eta^2 + \xi^2}, \quad \theta = \tan^{-1}\left(\frac{\xi}{\eta}\right),
\end{align}
where $(\eta,\xi)$ denote $(x,y)$, $(y,z)$, or $(x,z)$ axes, respectively. These boundary conditions generate a solution with features analogous to an L-shaped domain problem but comprising multiple edge and vertex singularities. While the exact solution for the problem is unknown, the convergence of the DPG residual is shown in~\Cref{resConvegefc}.

\begin{figure}[!htb]
\centering
\begin{subfigure}{0.45\textwidth}
	\includegraphics[width=\textwidth]{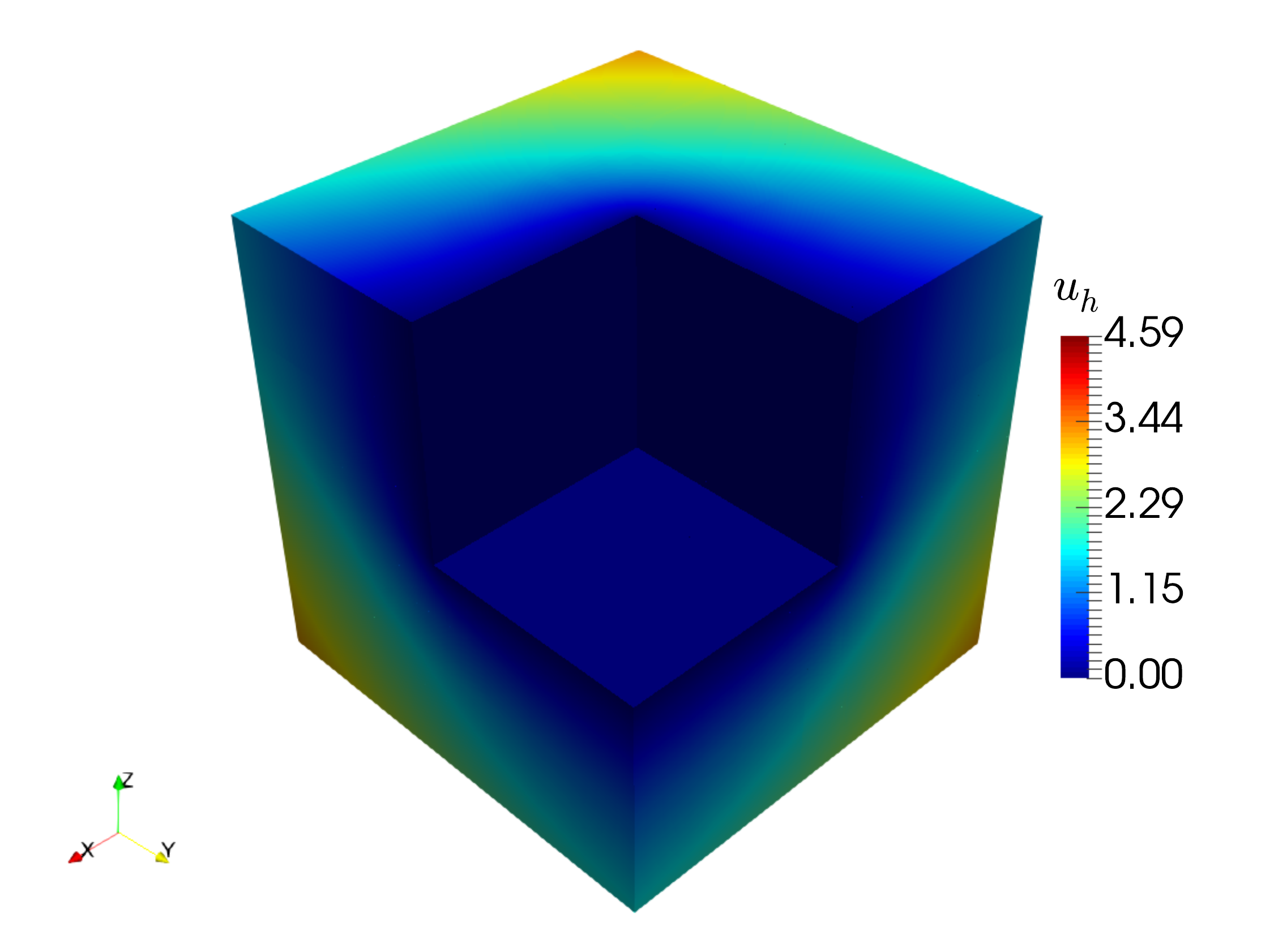}
	\caption{Isometric view along $(-1,-1,-1)$}
\end{subfigure}
\hspace{0.05\textwidth}
\begin{subfigure}{0.45\textwidth}
	\includegraphics[width=\textwidth]{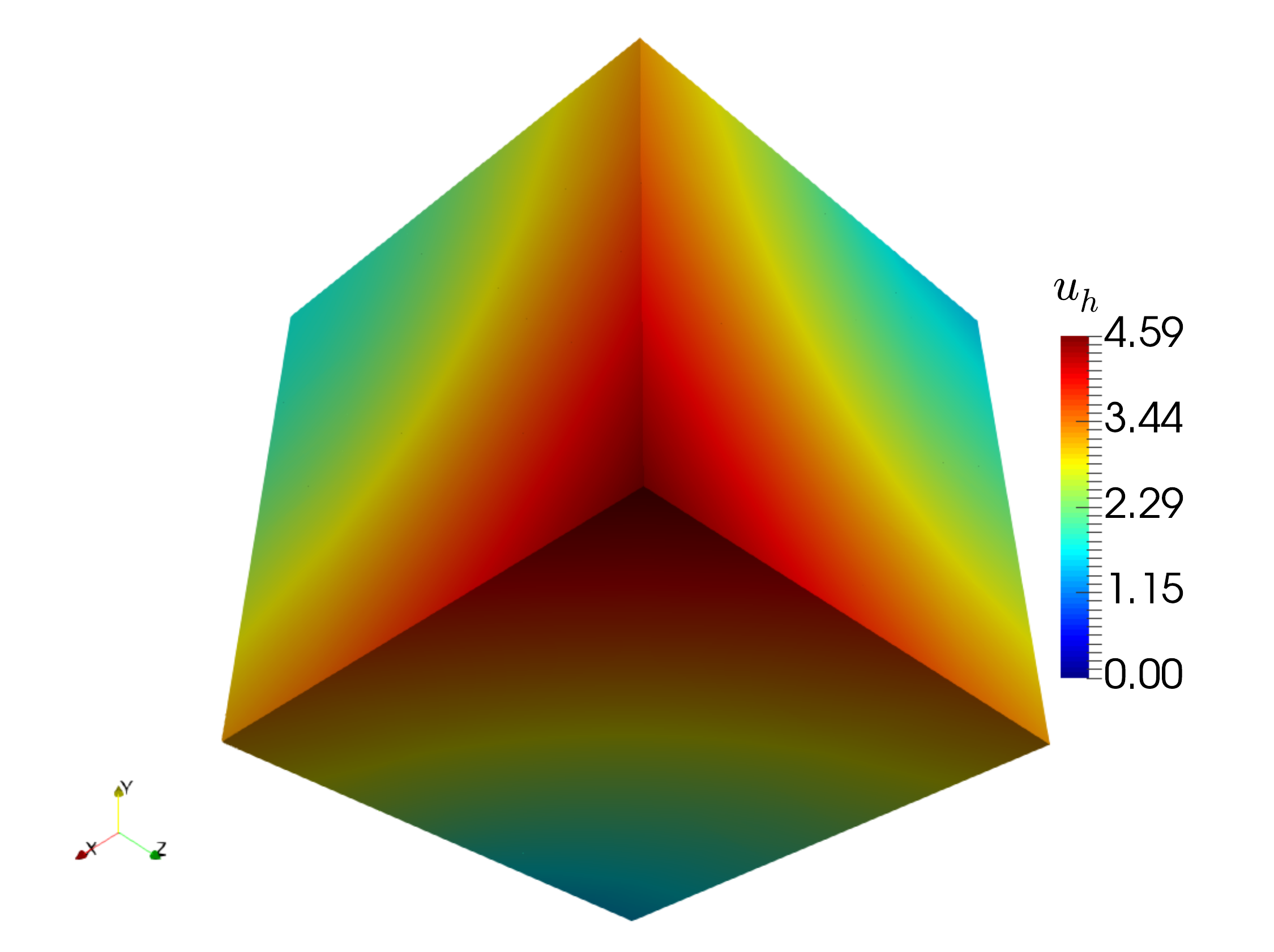}
	\caption{Isometric view along $(1,1,1)$}
\end{subfigure}
\caption{Fichera cube problem: solution contour. The problem is driven by the Neumann boundary conditions on the L-shaped faces in (a) and the three visible square faces in (b). The faces aligned along the coordinate planes in (a) have the Dirichlet boundary conditions.}\label{sol_FC}
\end{figure}

\begin{figure}[!htb]
\centering
\begin{subfigure}{0.45\textwidth}
	\includegraphics[width=\textwidth]{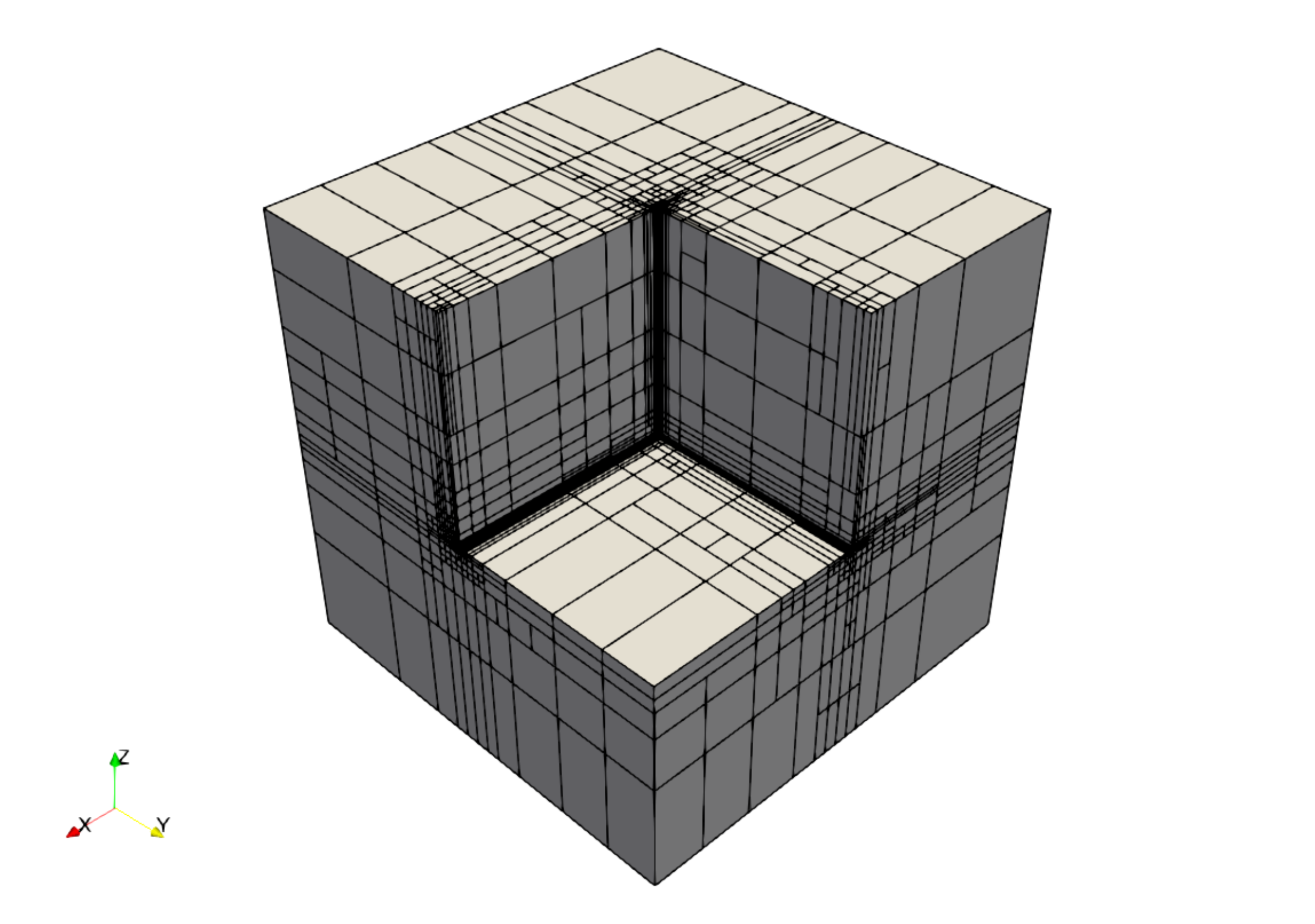}
	\caption{Isometric view along $(-1,-1,-1)$}
\end{subfigure}
\hspace{0.05\textwidth}
\begin{subfigure}{0.45\textwidth}
	\includegraphics[width=\textwidth]{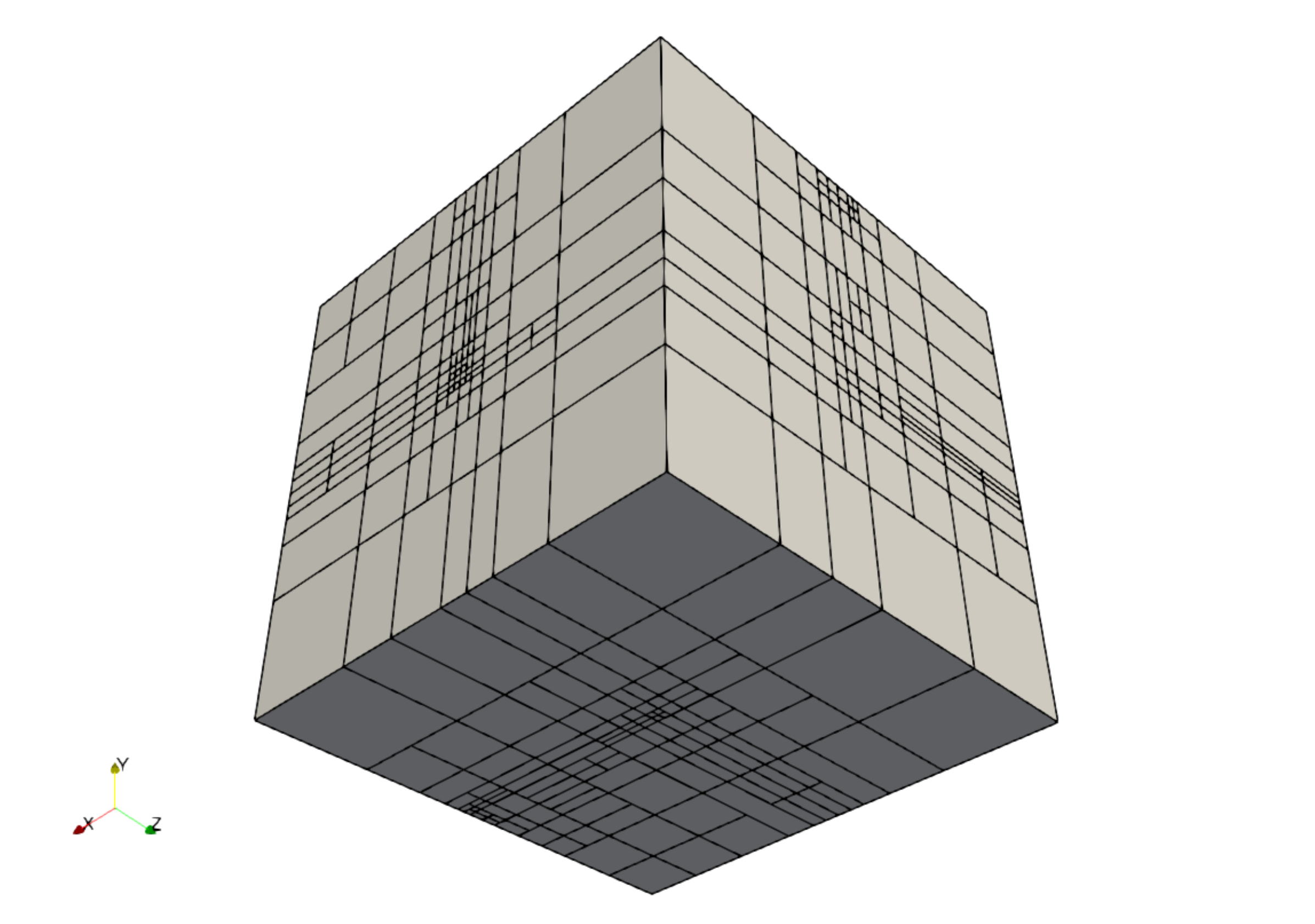}
	\caption{Isometric view along $(1,1,1)$}
\end{subfigure}
\caption{Fichera cube problem: an anisotropically adapted $hp$-mesh with 1.3M dofs.} \label{mesh_FC}
\end{figure}

\begin{figure}[!htb]
\centering
\begin{subfigure}{0.32\textwidth}
	\includegraphics[width=\textwidth]{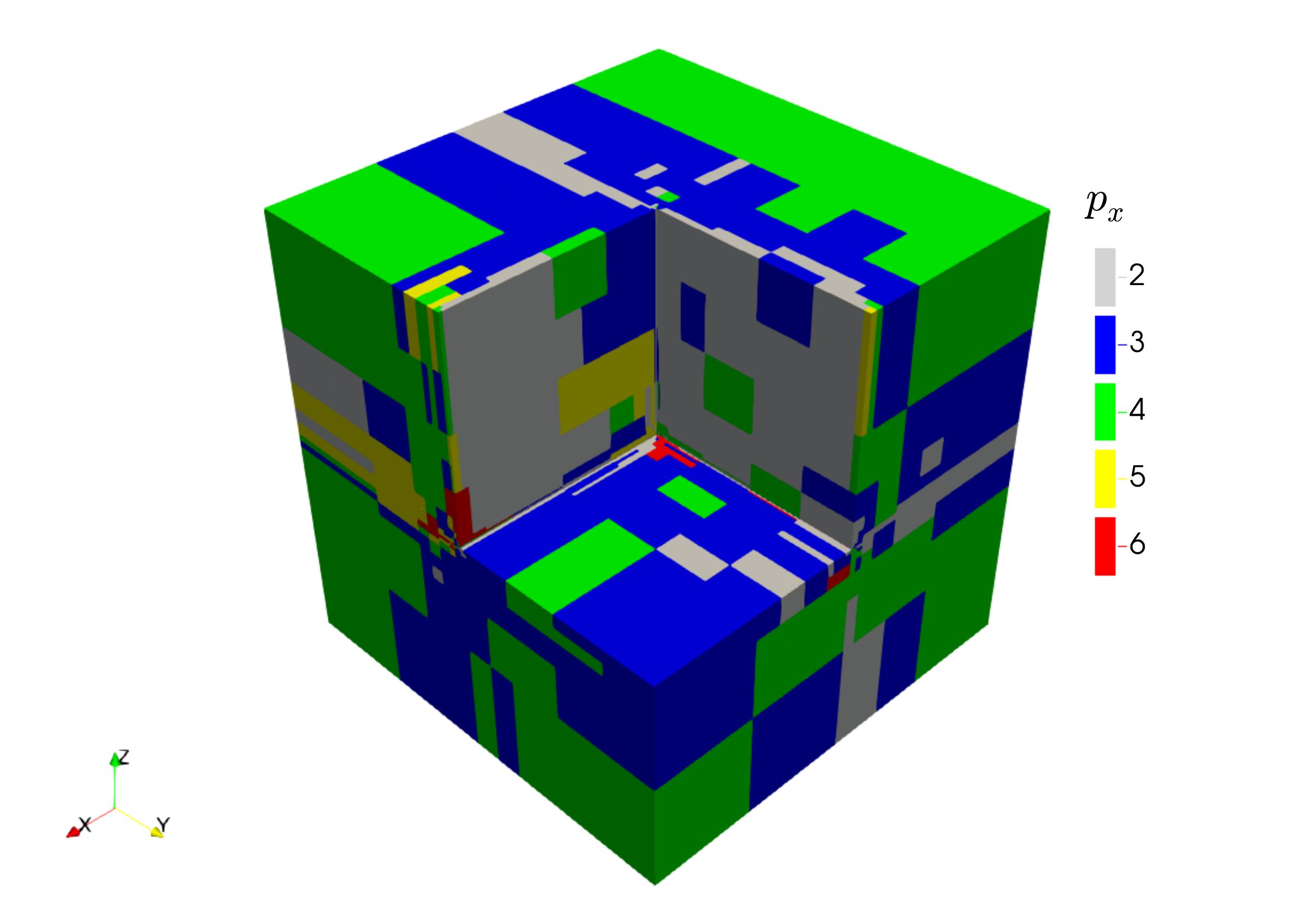}
	\caption{Polynomial order $p_x$}
\end{subfigure}
	\begin{subfigure}{0.32\textwidth}
	\includegraphics[width=\textwidth]{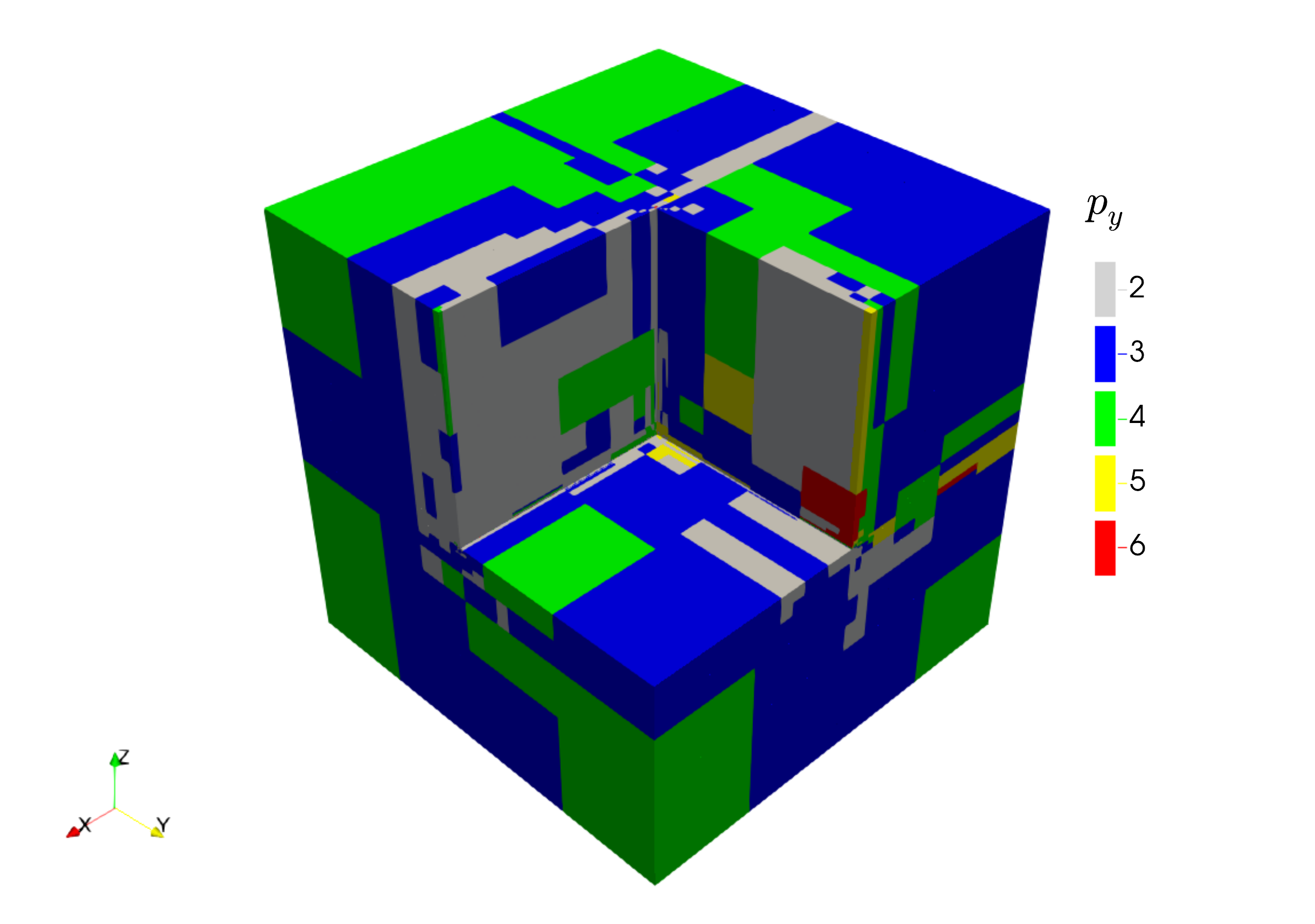}
	\caption{Polynomial order $p_y$}
\end{subfigure}
\begin{subfigure}{0.32\textwidth}
	\includegraphics[width=\textwidth]{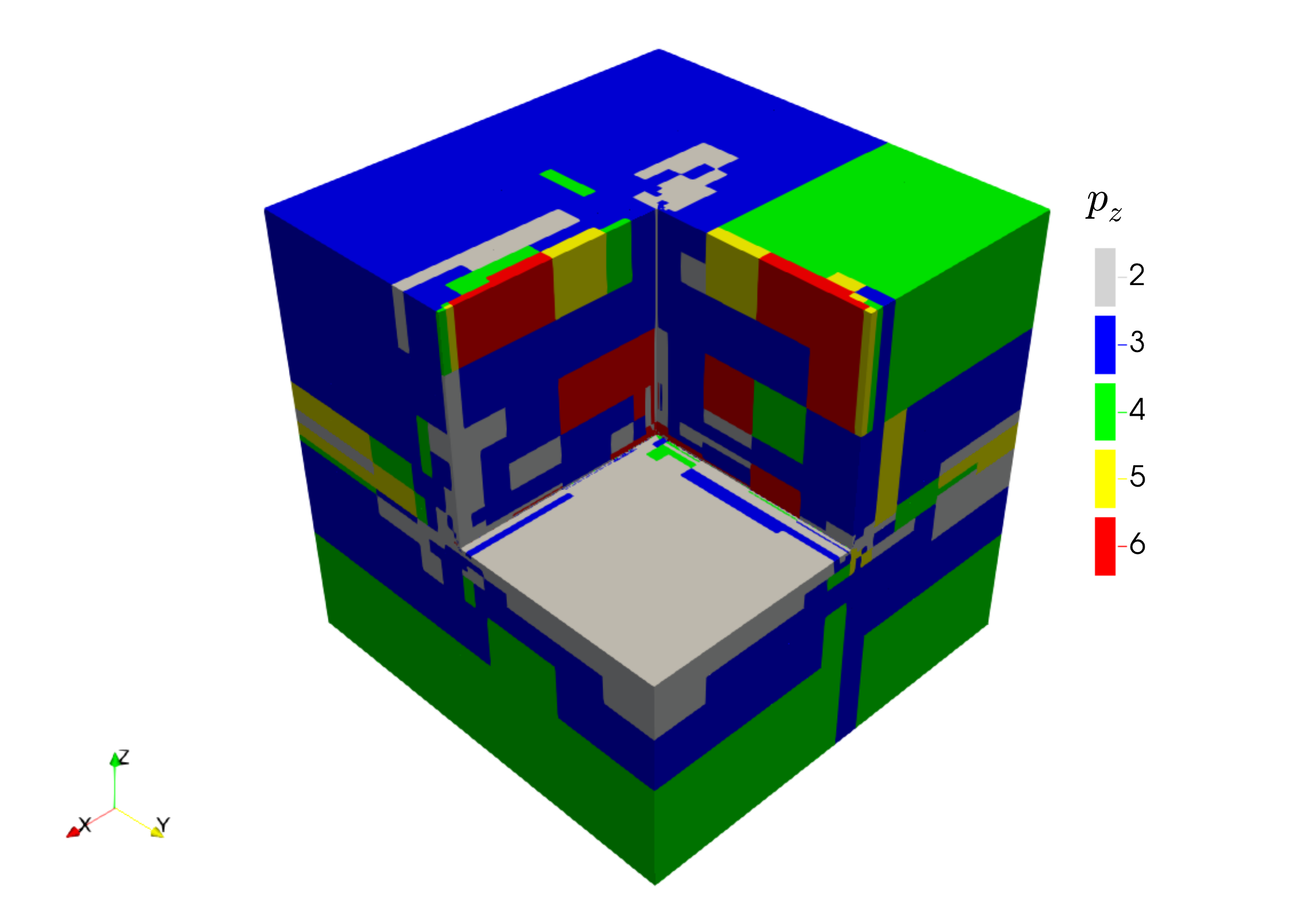}
	\caption{Polynomial order $p_z$}
\end{subfigure}
\caption{Fichera cube problem: polynomial distribution on the adapted $hp$-mesh. The algorithm prescribes low-order polynomials anisotropically around each edge singularity along $x$-, $y$- and $z$-axis. \Cref{poly_dist_FC_mgn} presents a magnified view of the polynomial distribution and anisotropic mesh elements around the singularities.} \label{poly_dist_FC}
\end{figure}

\begin{figure}[!htb]
\centering
\begin{subfigure}{0.45\textwidth}
	\includegraphics[width=\textwidth]{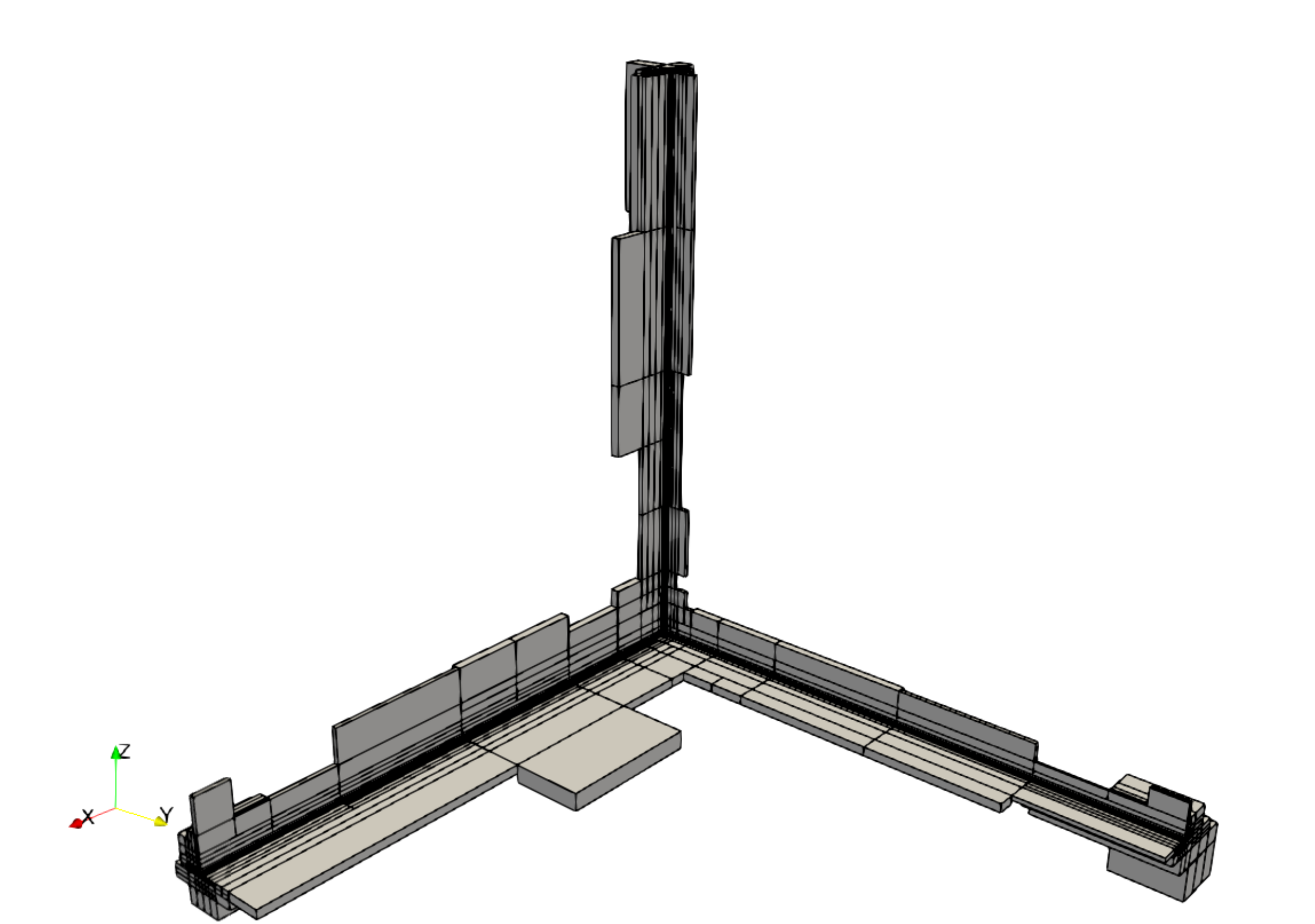}
	\caption{Adapted mesh}
\end{subfigure}
\hspace{0.05\textwidth}
\begin{subfigure}{0.45\textwidth}
	\includegraphics[width=\textwidth]{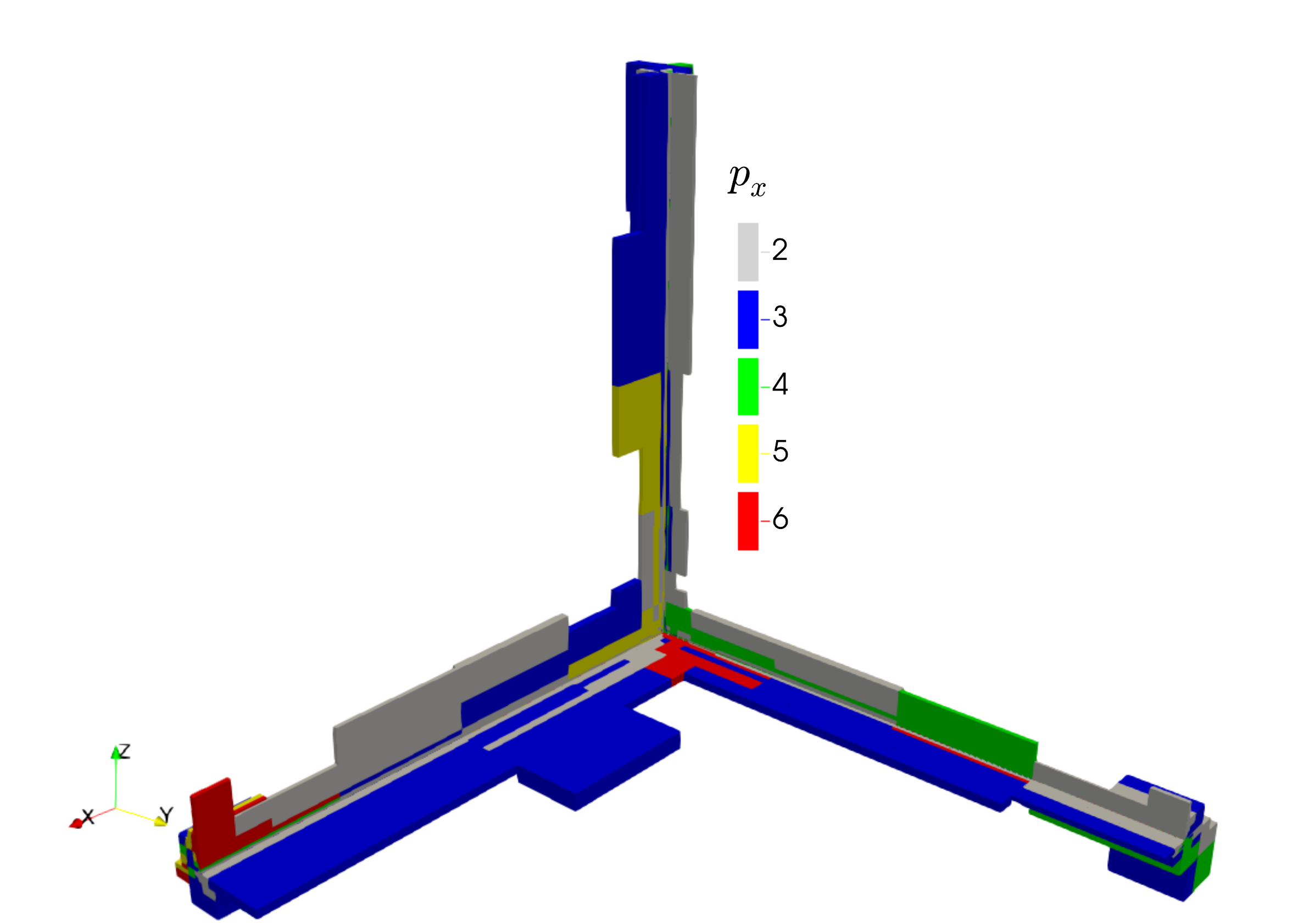}
	\caption{Polynomial order $p_x$}
\end{subfigure}

\begin{subfigure}{0.45\textwidth}
	\includegraphics[width=\textwidth]{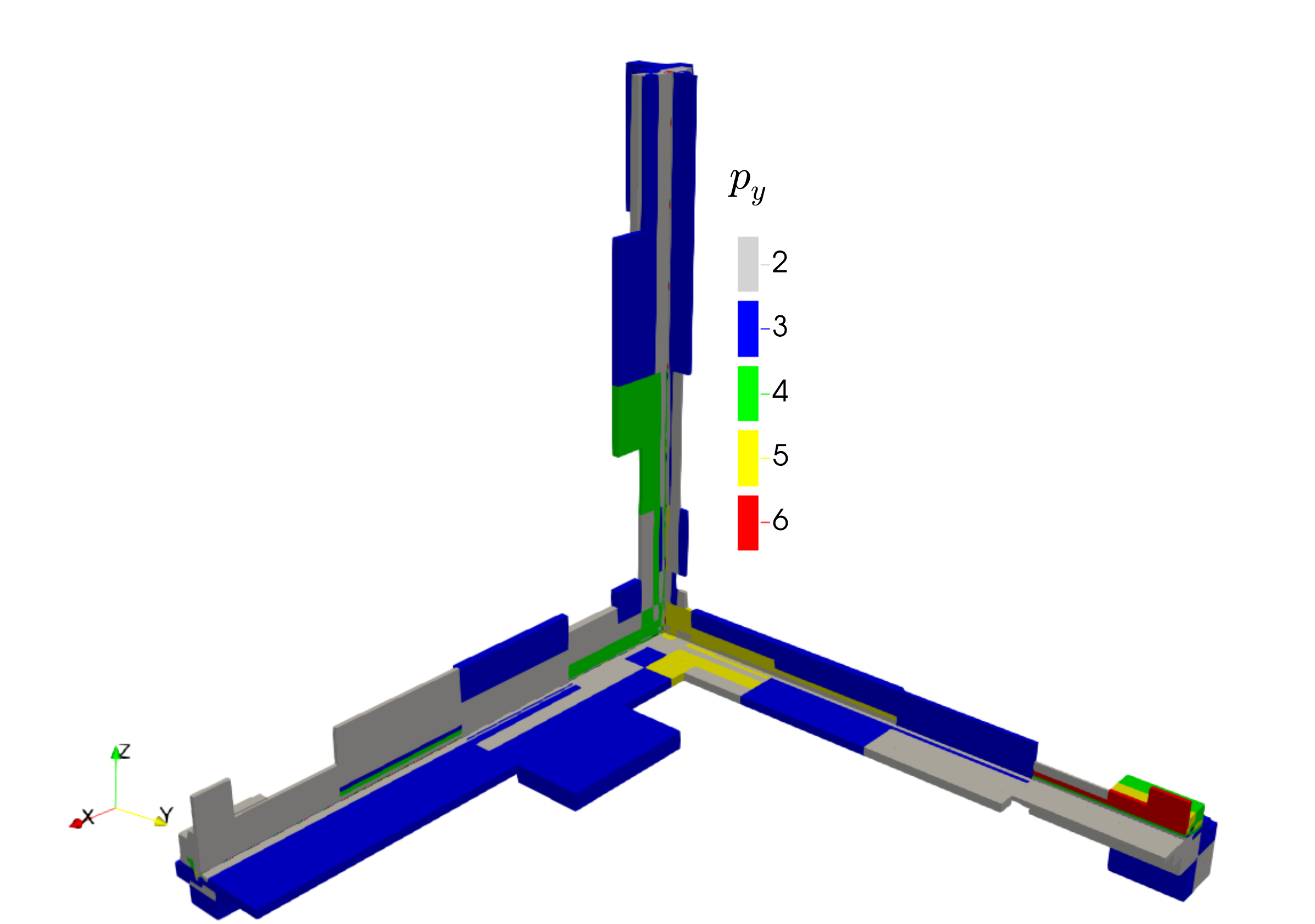}
	\caption{Polynomial order $p_y$}
\end{subfigure}
\hspace{0.05\textwidth}
\begin{subfigure}{0.45\textwidth}
	\centering
	\includegraphics[width=\textwidth]{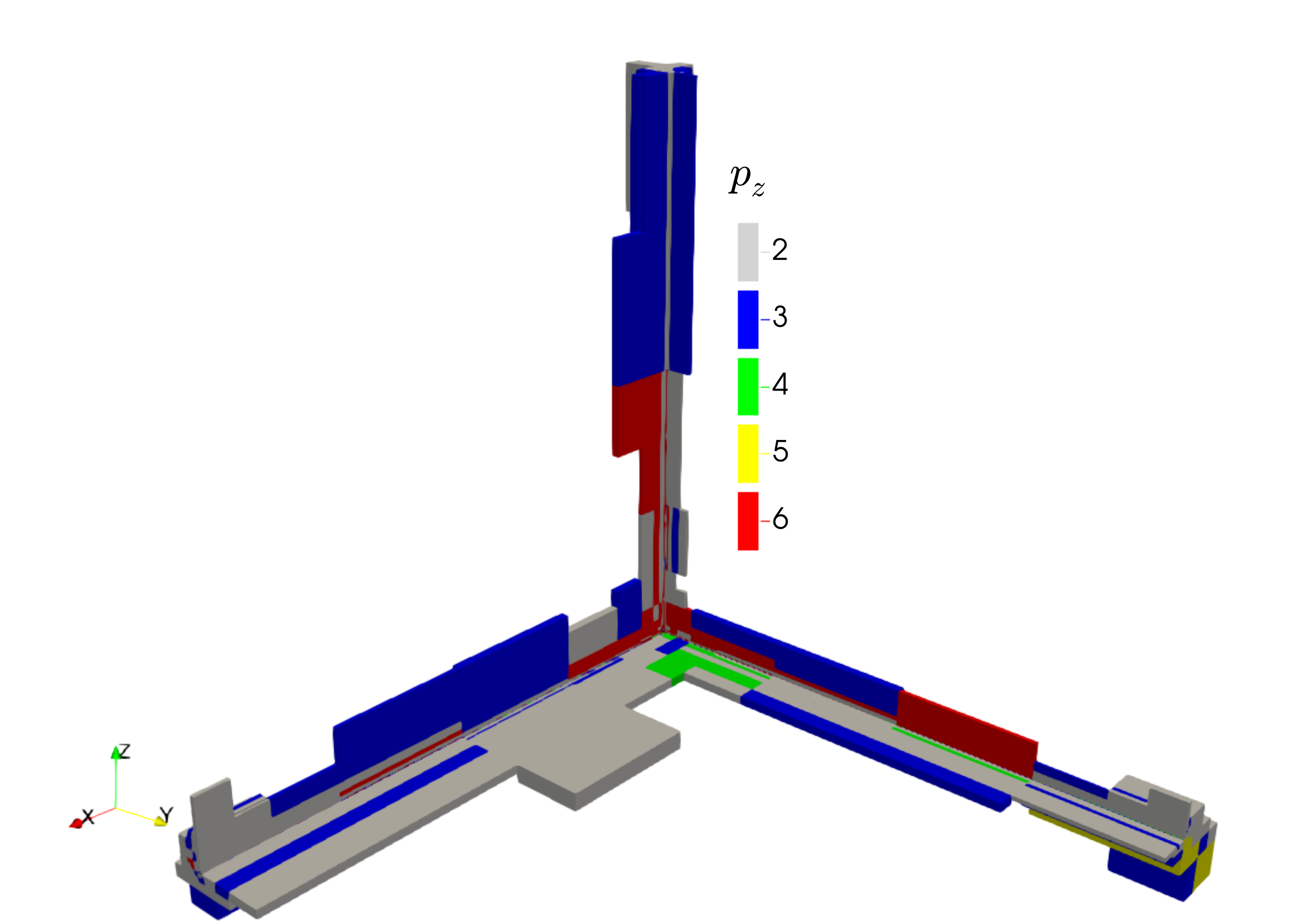}
	\caption{Polynomial order $p_z$}
\end{subfigure}

\caption{Fichera cube problem: magnified view of the mesh and the polynomial distribution near the edge and vertex singularities.} \label{poly_dist_FC_mgn}
\end{figure}

\Cref{sol_FC,mesh_FC} depict the solution contour and the corresponding adapted mesh, respectively. \Cref{poly_dist_FC,poly_dist_FC_mgn} illustrate the polynomial distribution associated with the adapted mesh.~\Cref{mesh_FC} shows that the refinement algorithm performs highly anisotropic $h$-refinements along the edge singularities, generating graded meshes. The anisotropic refinements propagate through the volume to the opposing boundary faces on $\Gamma_\sigma$. The propagation of refinements happens in conjunction to the singularities arising from the faces with Neumann boundary conditions. \Cref{poly_dist_FC,poly_dist_FC_mgn} clearly indicate that the algorithm chooses lowest order polynomials around the singularities. Moving away from the singularities, the algorithm prescribes higher order polynomials underscoring the smoothness of the solution variables. In~\Cref{resConvegefc}, one can observe the exponential convergence of the residual on performing $hp$-refinements, whereas isotropic $h$-refinements suffer from a loss of convergence due to the lack of required grading in size and polynomial distribution.
 
\subsection{Eriksson--Johnson Problem}
We consider a convection-dominated diffusion problem motivated by the Eriksson--Johnson model problem \cite{erjk}. Here, we extend the exact solution of the two-dimensional problem by multiplying it with a sinusoidal term along $z$. In particular, we solve
\begin{align}
\begin{split}
\frac{\partial u}{\partial x} - \epsilon \nabla^2 u &= f(x,y,z)  \qquad \qquad \qquad \, \text{in} \quad \Omega:={(0,1)}^3 , \\
u &= 0 \qquad \qquad \qquad \qquad \quad \, \, \, \text{on} \quad \Gamma_{u_a},\\
u &= \sin(\pi y) \sin(\pi z)   \qquad  \quad \, \, \, \text{on} \quad \Gamma_{u_b},
\end{split}
\end{align}
where
\be
\begin{split}
\Gamma_{u_a} = \partial \Omega \setminus \{0\} \times [0,1] \times [0,1] 
\quad \text{and} \quad
\Gamma_{u_b} =  \{0\} \times [0,1] \times [0,1].
\end{split}
\ee

The source $f$ and the boundary conditions are computed using the exact solution given by
\be
	u(x,y,z) = \frac{e^{s_1(x-1)} - e^{s_2(x-1)}}{e^{s_1} - e^{s_2}} \sin(\pi y) \sin(\pi z),
\ee
where
\be
	s_1 = \frac{1 + \sqrt{1 + 4 \pi^2 \epsilon^2}}{2\epsilon} 
	\text{ and } 
	s_2 = \frac{1 - \sqrt{1 + 4 \pi^2 \epsilon^2}}{2\epsilon} .
\ee

\begin{figure}[!htb]
\centering
\begin{subfigure}{0.45\textwidth}
	\includegraphics[width=\textwidth]{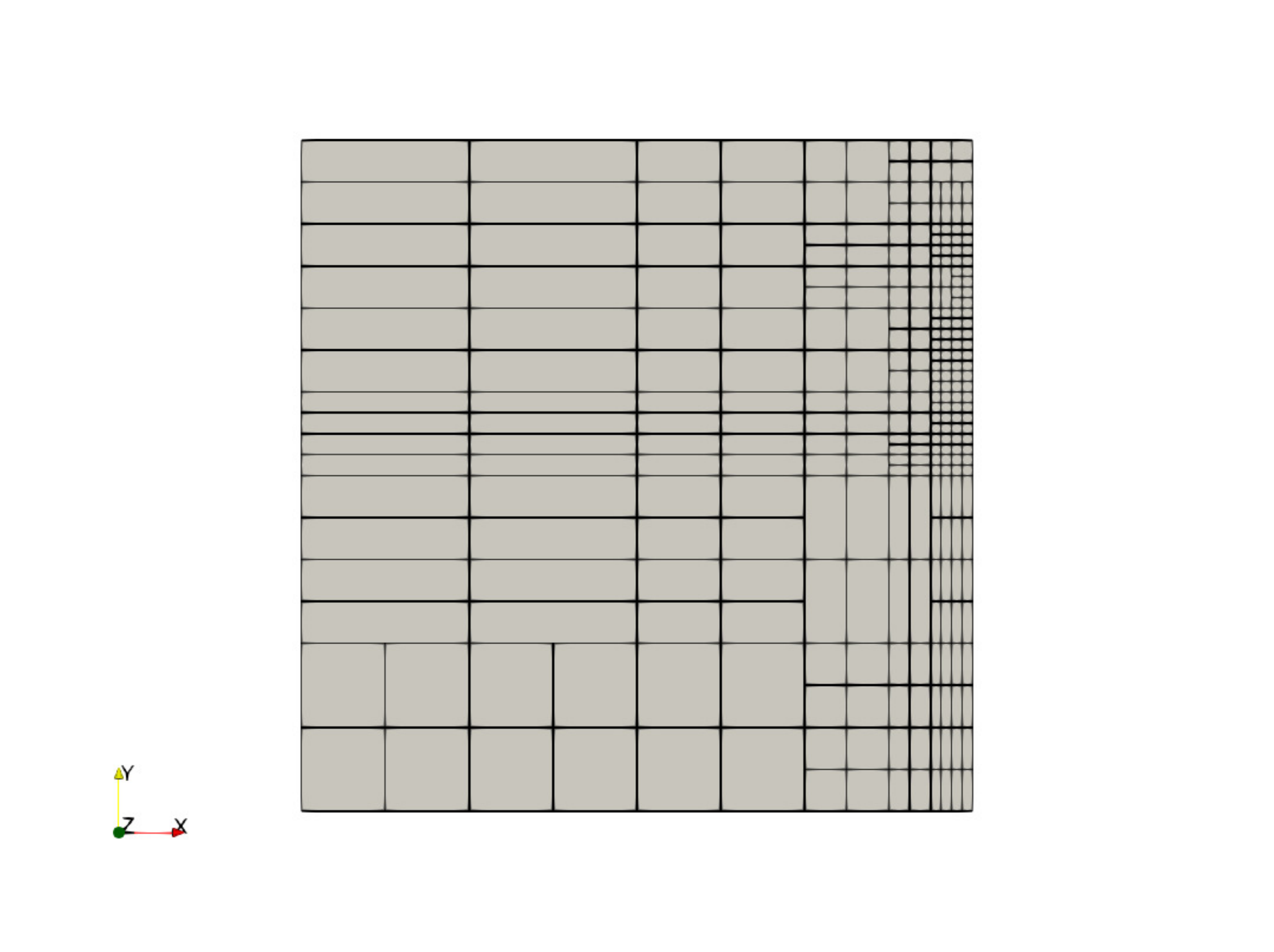}
	\caption{Cross-section of an adapted mesh at $z = 0.5$}
\end{subfigure}
\hspace{0.05\textwidth}
\begin{subfigure}{0.45\textwidth}
	\includegraphics[width=\textwidth]{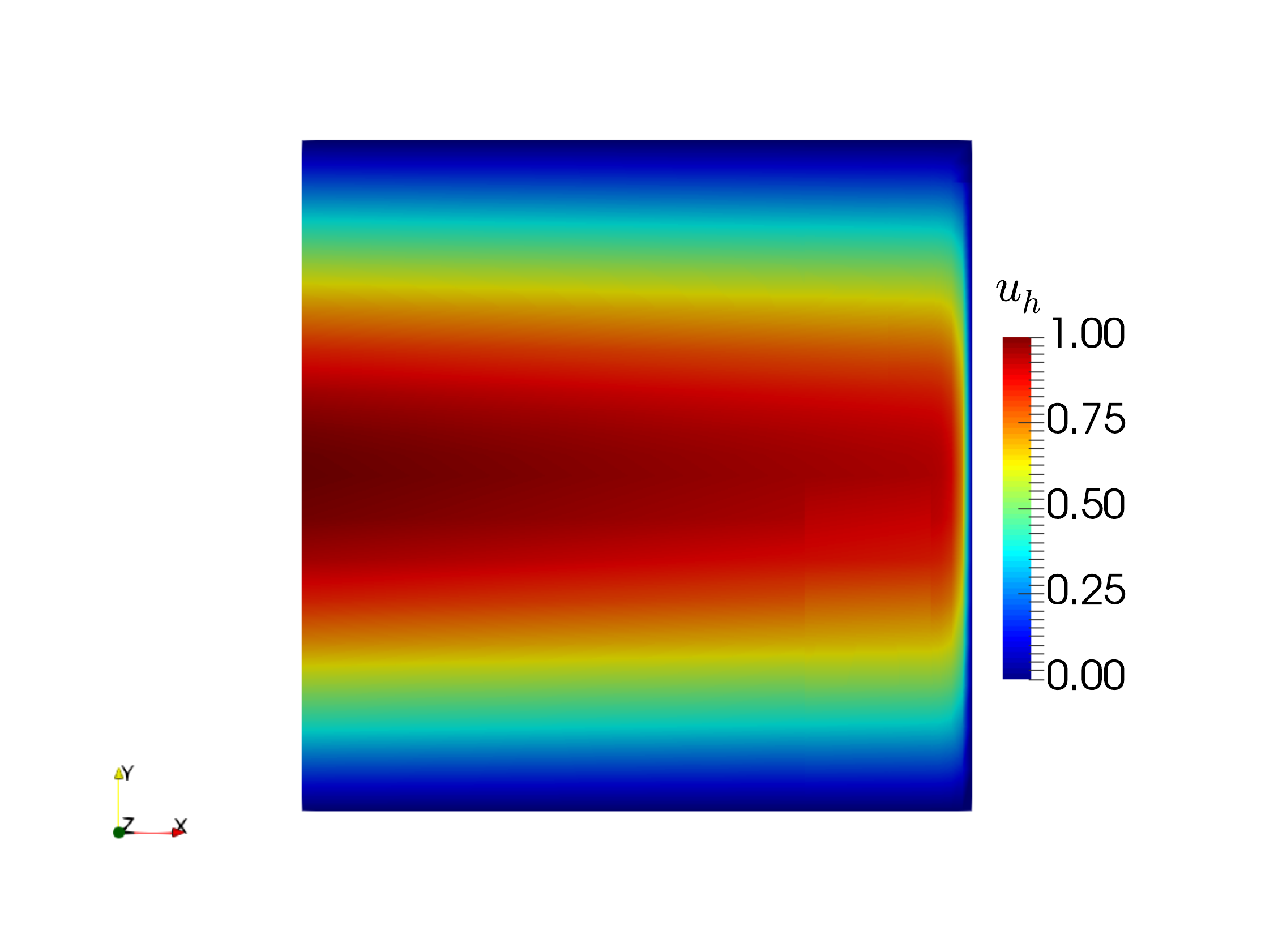}
	\caption{Solution contour at $z = 0.5$}
\end{subfigure}
\caption{Eriksson--Johnson problem: an adapted mesh and solution contour.} \label{EJ_sol}
\end{figure}

\begin{figure}[!htb]
\centering
\begin{subfigure}{0.45\textwidth}
	\includegraphics[width=\textwidth]{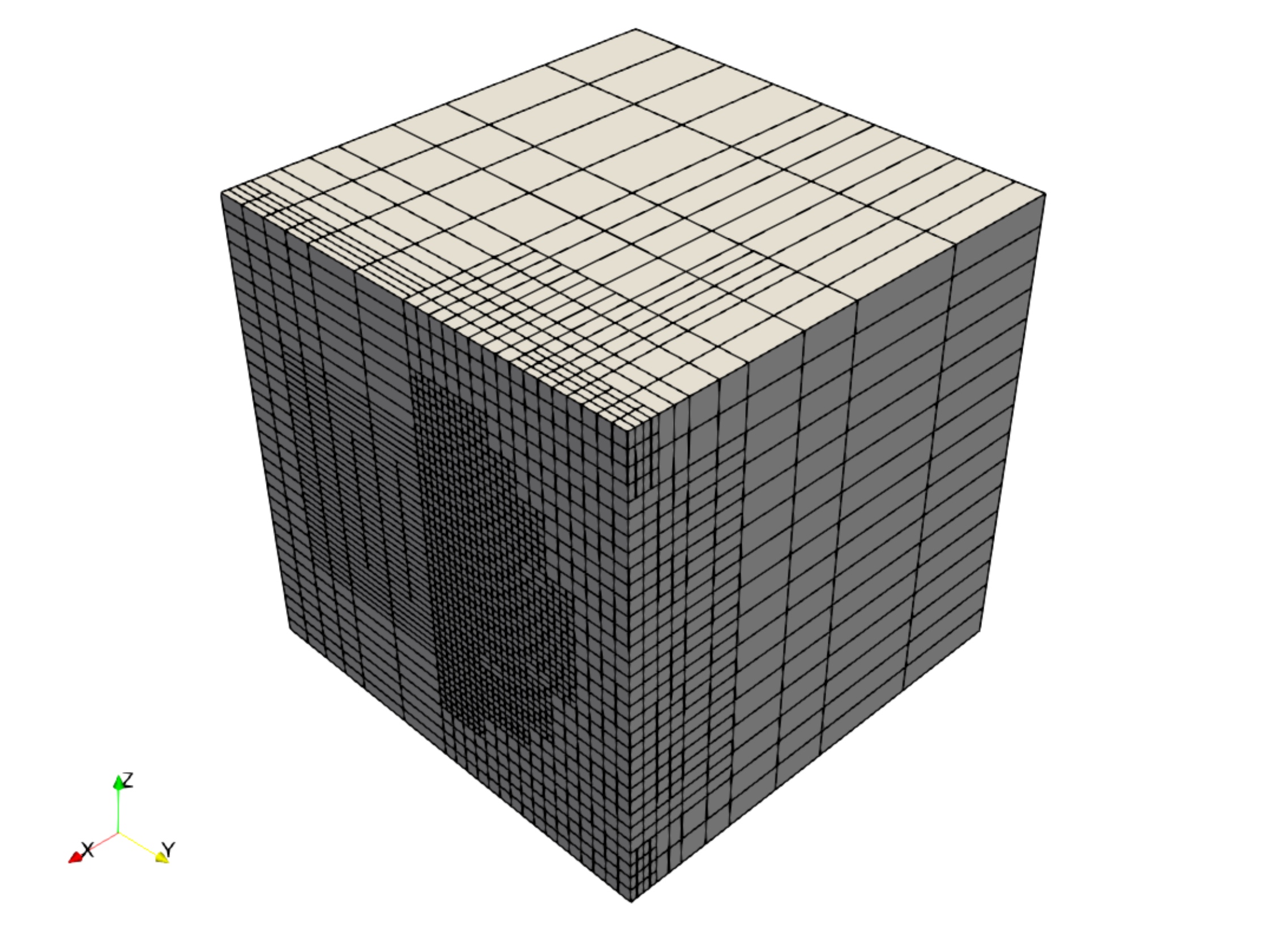}
	\caption{Isometric view of the mesh.}
\end{subfigure}
\hspace{0.05\textwidth}
\begin{subfigure}{0.45\textwidth}
	\includegraphics[width=\textwidth]{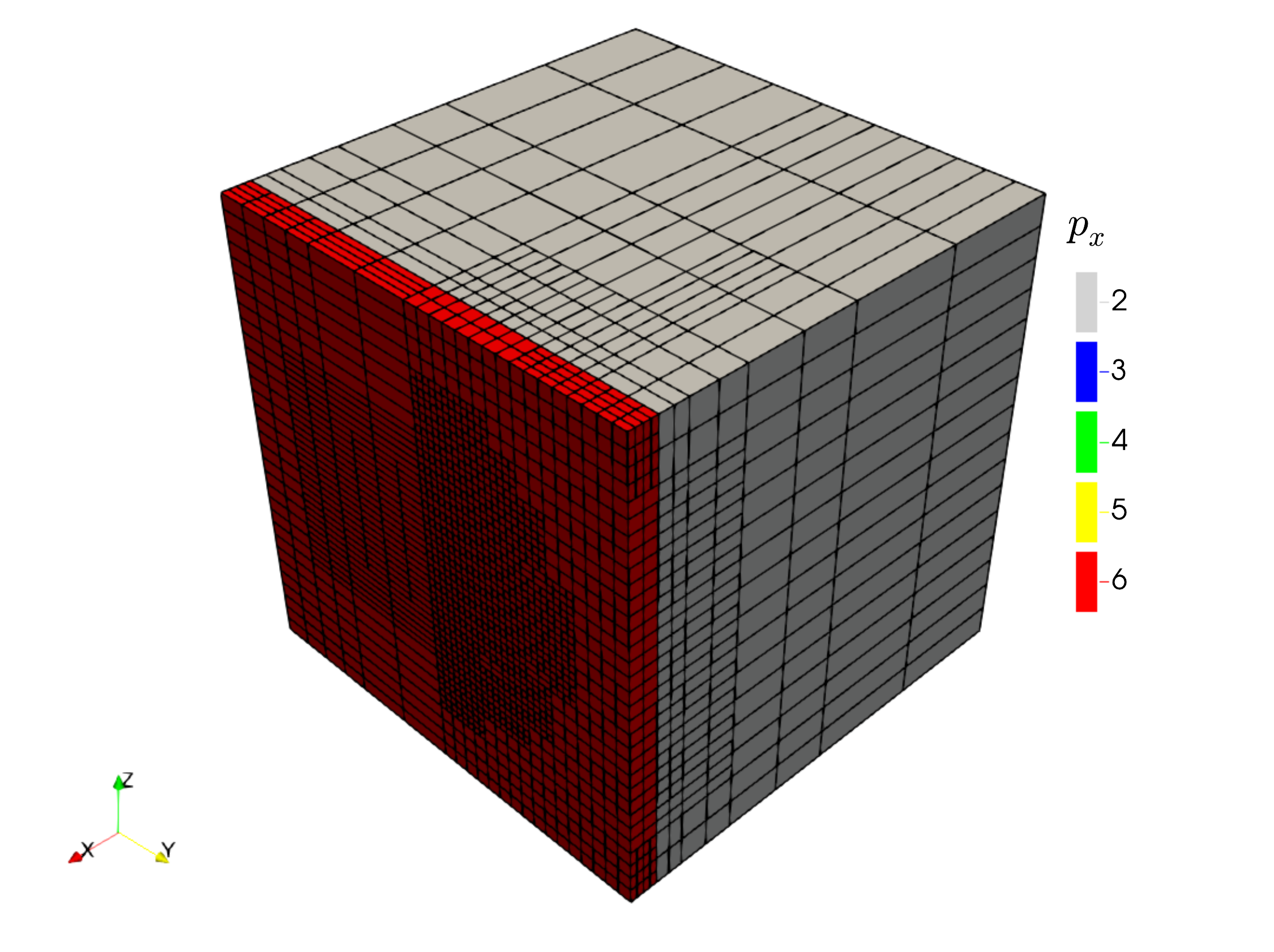}
	\caption{Polynomial order along $x$ direction: $p_x$}
\end{subfigure}
\caption{Eriksson--Johnson problem: an adapted mesh with $209\,737$ dofs; coloring indicates the corresponding polynomial distribution along the $x$-axis.} \label{poly_dist_EJ}
\end{figure}

\begin{figure}[!htb]
\centering
\resizebox{0.45\textwidth}{!}{
\begin{tikzpicture}
		\begin{semilogyaxis}[xmin=6,xmax=200, ymin=1e-4,ymax=1,xlabel=\large{$\sqrt[3]{\text{ndof}}$},ylabel=\large{$\text{Relative } L^2 \text{ error}$},grid=major,legend style={at={(1,1)},anchor=north east,font=\small,rounded corners=2pt}]
		\addplot[color = blue,mark=square*] table[x= ndof_tot, y=err_L2_rel, col sep = comma] {results/EJ_testcase_invst_adjoint_graph/error_p2_EJ_adj_results.txt};
		\addplot [color = red,mark=square*] table[x= ndof_tot, y=err_L2_rel, col sep = comma] {results/EJ_testcase_invst_adjoint_graph/error_p3_EJ_adj_results.txt};
		\addplot [color = black,mark=square*] table[x= ndof_tot, y=err_L2_rel, col sep = comma] {results/EJ_testcase_invst_adjoint_graph/error_p4_EJ_adj_results.txt};
		\addplot [color = magenta,mark=square*] table[x= ndof_tot, y=err_L2_rel, col sep = comma] {results/EJ_testcase_invst_adjoint_graph/error_hp_EJ_adj_invst_results.txt};
		\legend{$p = 2$,$p = 3$,$p = 4$,$hp$}
		\end{semilogyaxis}
\end{tikzpicture}
}
\hspace{0.05\textwidth}
\resizebox{0.45\textwidth}{!}{
\begin{tikzpicture}
		\begin{semilogyaxis}[xmin=6,xmax=200, ymin=1e-4,ymax=1,xlabel=\large{$\sqrt[3]{\text{ndof}}$},ylabel=\large{$\text{Residual}$},grid=major,legend style={at={(1,1)},anchor=north east,font=\small,rounded corners=2pt} ]
		\addplot[color = blue,mark=square*] table[x= ndof_tot, y=res, col sep = comma] {results/EJ_testcase_invst_adjoint_graph/error_p2_EJ_adj_results.txt};
		\addplot [color = red,mark=square*] table[x= ndof_tot, y=res, col sep = comma] {results/EJ_testcase_invst_adjoint_graph/error_p3_EJ_adj_results.txt};
		\addplot [color = black,mark=square*] table[x= ndof_tot, y=res, col sep = comma] {results/EJ_testcase_invst_adjoint_graph/error_p4_EJ_adj_results.txt};
		\addplot [color = magenta,mark=square*] table[x= ndof_tot,  y=res, col sep = comma] {results/EJ_testcase_invst_adjoint_graph/error_hp_EJ_adj_invst_results.txt};
		\legend{$p = 2$,$p = 3$,$p = 4$,$hp$}
		\end{semilogyaxis}
\end{tikzpicture}
}
\caption{Eriksson--Johnson problem: convergence of relative $L^2$ error and DPG residual.}\label{resConvegeEj}
\end{figure}
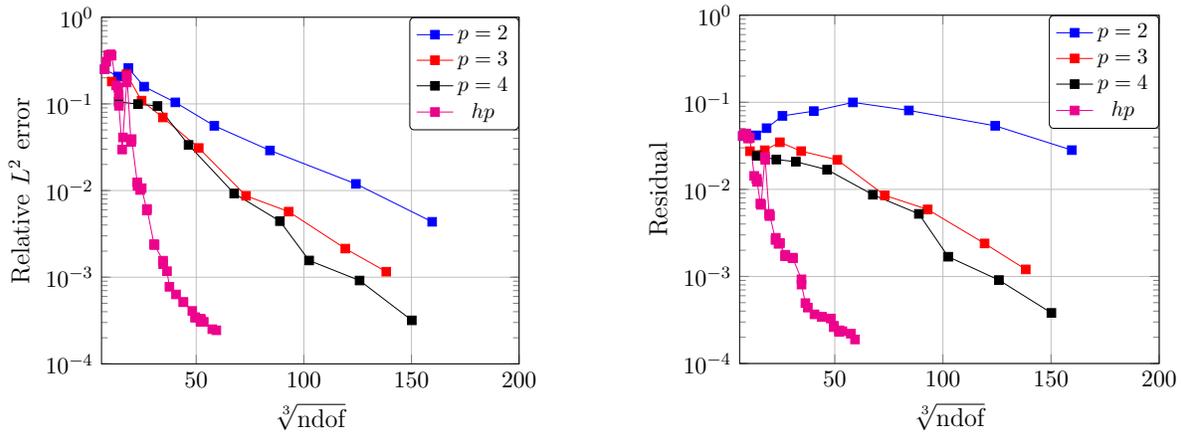

In this numerical experiments, $\epsilon = 0.01$. \Cref{EJ_sol} depicts the cross-section of an adapted mesh and the corresponding solution contour at $z=0.5$. The solution exhibits a boundary layer along the $x$-axis with sinusoidal variations along $y$ and $z$. The variation in the solution is also reflected in the $hp$-refinements executed by the algorithm. In order to capture the boundary layer, the algorithm generates anisotropic elements parallel to the $yz$-plane and assigns the highest polynomial order along the $x$-axis inside the boundary layer. Since the boundary layer is weighted with sinusoidal variations in $y$ and $z$, the majority of the $h$-refined elements in the boundary layer are positioned near $y = 0.5$ and $z=0.5$. \Cref{poly_dist_EJ} illustrates the adapted mesh with the polynomial distribution along the $x$-axis. Finally, \Cref{resConvegeEj} presents the convergence plots for the relative $L^2$ error and the residual, demonstrating the efficacy of the proposed $hp$-refinement strategy for this problem.

%
%

\section{Conclusion
\label{sec:conclusions}
} 
The anisotropic $hp$-refinement strategy presented in this article utilizes the built-in DPG error-estimator and $L^2$ projection-based error estimates for the ultraweak variational formulation. The efficacy of the proposed algorithm is demonstrated through numerical experiments containing boundary layers and singularities. 
The algorithm is able to generate a sequence of meshes that provide exponential convergence. Since we have capped the maximum polynomial order in our numerical experiments to  $p = 6$ for practical reasons, we observe a slight loss of optimal convergence rate. Nonetheless, the accuracy of the solutions on the anisotropically refined $hp$-meshes remains orders of magnitude better than that on isotropically refined meshes for nearly same number of dof. The proposed $hp$-refinement strategy complements anisotropic $h$-refinements with anisotropic $p$-refinements, which allows the algorithm to avoid any superfluous investment (in terms of dofs).

\paragraph{Future work} To accelerate the computation of the fine-grid solution and apply the $hp$-refinement strategy to large-scale multiphysics problems, we intend to integrate the proposed $hp$-refinement strategy with the scalable DPG-MG solver \cite{badgerMG}. Additionally, we aim to extend the proposed refinement strategy to other element types, such as tets, prisms, and pyramids, in order to leverage $hp$3D's capability to handle hybrid meshes.

\printbibliography

\end{document}